\documentclass[a4paper]{aa}
\usepackage{graphicx}
\usepackage{txfonts}
\usepackage{flafter}
\usepackage{fixltx2e}

\bibpunct{(}{)}{;}{a}{}{,} 

\begin{document}
\title{A connected component-based method for efficiently integrating multiscale $N$-body systems}
\titlerunning{A connected components-based $N$-body integrator}
\author{
J\"urgen J\"anes\inst{1}
\thanks{Current address: The Gurdon Institute and Department of Genetics, University of Cambridge, Cambridge CB3 0DH, United Kingdom} 
\and Inti Pelupessy\inst{2}
\thanks{ Corresponding author. Tel.: +31 715278424; fax: +31 715275743. E-mail address: pelupes@strw.leidenuniv.nl (F.I. Pelupessy).}
\and Simon Portegies Zwart\inst{2}
}
\authorrunning{J. J\"anes et al.}

\institute{
Faculty of Science, University of Amsterdam, P.O. Box 94216, 1090 GE, Amsterdam, The Netherlands\and
Leiden Observatory, Leiden University, P.O. Box 9513, 2300 RA, Leiden, The Netherlands
}

\date{Received XX; accepted YY}

\abstract{
We present a novel method for efficient direct integration of gravitational N-body systems with a large variation in characteristic time scales. The method is based on a recursive and adaptive partitioning of the system based on the connected components of the graph generated by the particle distribution combined with an interaction-specific time step criterion. It uses an explicit and approximately time-symmetric time step criterion, and conserves linear and angular momentum to machine precision. In numerical tests on astrophysically relevant setups, the method compares favourably to  both alternative Hamiltonian-splitting integrators 
 as well as recently developed block time step-based GPU-accelerated Hermite codes.
Our reference implementation is incorporated in the HUAYNO code, which is freely available as a part of the AMUSE framework. 
}


\maketitle

\section{Introduction}

Direct integration of the classical $N$-body problem is an important tool for studying astrophysical systems. Examples include planetary systems, open and globular clusters dynamics, large-scale dynamics of galaxies, and structure formation in the universe. In many cases the calculations involve systems where the intensity of gravitational interactions spans multiple orders of magnitude with corresponding timescale variations. For example, the initial stages of cluster formation are now thought to resemble multi-scale fractal structures \citep{GoodwinWhitworth2004}, and stellar systems are invariably formed with a high fraction of binaries and hierarchical multiples that affect the dynamical evolution in crucial ways \citep{PortegiesZwart2010}.

In practice, integrating multiscale systems requires specialised methods that vary the resolution at which we treat different parts of the simulation. The aim of this is to obtain a solution with an acceptable accuracy without unnecessarily spending computational resources on the slowly evolving parts of the simulation. In generic $N$-body integrators, this idea is most commonly implemented via {\em particle-based block time steps} --- every particle in the system maintains an individual time step limited to discrete values in a power of two hierarchy. These block time steps are then typically used to determine the frequency of calculating the total force acting on a particle \citep[e.g.][]{McMillan1986, Makino1991,Konstantinidis:2010hx}.

While considerably speeding up calculations, particle-based block time steps are nevertheless limited in their ability to treat the extreme scale differences often present in $N$-body systems. Hence, complementary strategies, such as binary regularisation and neighbourhood schemes, have been devised. These approaches complicate the implementation of $N$-body integrators, and often introduce new method-specific free parameters. It is also unclear whether these combinations of multiple strategies represent the best possible approach for integrating multiscale $N$-body systems. These issues provide a clear incentive to explore alternative methods.

In \cite{Pelupessy:2012if}, we derived generic $N$-body integrators that recursively and adaptively split the Hamiltonian of the system. These methods show improved conservation of the integrals of motion by always evaluating partial forces between particles in different time-step bins symmetrically, and by using an approximately time-symmetric time-step criterion. 

In the present work, we introduce a new Hamiltonian-splitting integration method that is particularly adept at integrating initial conditions with significant hierarchical substructure. Our approach is based on assigning time steps to individual interactions, followed by partitioning the system Hamiltonian based on a graph formed by the set of interactions that are faster than a fixed threshold time step. The successive partitioning produces closed Hamiltonians such that we can easily use specialised solvers for situations where more efficient solvers are available. Numerical experiments show that our integrator compares favourably to existing methods even for an ordinary Plummer sphere where the prevalence of isolated subsystems is not immediately obvious. For astrophysically realistic systems explicitly chosen for their multi-scale substructure, the performance gains increase can be orders of magnitude. An implementation of the method is incorporated in the HUAYNO code, which is freely available as a part of the AMUSE framework\citep{PortegiesZwart2013b, Pelupessy2013Amuse} and which was used for the tests presented in this paper.

Our method is similar in spirit, and accelerates the calculation of the N-body problem for much the same reasons as the well known neighbour schemes. The main idea is to divide the total force acting on a particle into a fast and a slow component based the distance to the given particle.  Different approaches have been used for treating fast and slow components. The Ahmad-Cohen neighbourhood scheme\citep{Ahmad:1973kn} treats fast components with a more strict time step criteria. Alternatively, the PPPT scheme\citep{Oshino:2011wv} integrates fast components with a fourth-order Hermite method while using a leapfrog-based tree code for the long range interactions. The criteria for determining neighbourhood memberships are heuristics known to work in numerical experiments, e.g. a sphere with a fixed radius centred on the acting particle. These methods need to continuously update neighbourhood memberships as the system state changes throughout the simulation. In addition, neighbourhood schemes only make a single distinction between treating small subsystems such as hard binaries or many-body close encounters, and the large scale dynamics. It is difficult to generalise a neighbourhood scheme beyond a binary differentiation of the particle distribution.

In Section \ref{sec:method} we describe a bottleneck in existing general $N$-body splitting methods, and derive our novel splitting scheme that overcomes this bottleneck. Section \ref{sec:tests} presents the results of numerical tests comparing of our method to existing approaches. Finally, in Section \ref{sec:discussion} we discuss possible improvements and extensions to our work, including the feasibility of integrating general $N$-body systems using purely interaction-specific time steps.

\section{Method}\label{sec:method}

\subsection{Deriving time stepping schemes via Hamiltonian splitting}\label{sec:splitting-schemes}

The Hamiltonian for a system of $N$ particles $i=1\ldots N$ under gravitational interaction can be represented as a sum of momentum terms $T_{i}$ and potential terms $V_{ij}$:
\begin{eqnarray}
\label{eq:Hamiltonian}
H(\mathbf{p}_{i},\mathbf{q}_{i}) & = & T+V=\underset{i=1}{\overset{N}{\sum}}T_{i}+\underset{\substack{i,j=1\\i<j}}{\overset{N}{\sum}}V_{ij}\\
T_{i} & = & \frac{\left|\mathbf{p}_{i}\right|^{2}}{2m_{i}}\\
V_{ij} & = & -G\frac{m_{i}m_{j}}{\sqrt{q_{ij}^{2}+\varepsilon^{2}}}
\end{eqnarray}
where $m_{i}$ is the mass, $\mathbf{q}_{i}$ is the position and $\mathbf{p}_{i}$ is the momentum of the $i$-th particle of the system, and $q_{ij} = \|\mathbf{q}_{i}-\mathbf{q}_{j} \|$ . The evolution of the state of the system for a time step $h$ is given formally by the flow operator $E_{H}(h) = \exp( h \mathbb{H})$ where $\mathbb{H}$ is the Hamiltonian vector field corresponding to $H$.

If the Hamiltonian $H$ of the system is representable as a sum of two sub-Hamiltonians, $H=A+B$, we can approximate the time evolution under $H$ with a sequence of time evolution steps under the sub-Hamiltonians $A$ and $B$. A straightforward successive application of the time evolution under $A$ followed by the time evolution under $B$ gives a first-order approximation of the full time evolution under $A+B$, while a second-order accurate approximation can be obtained with one additional operator evaluation (\cite{SSC94}, Sec 12.4, also \cite{Hairer2006}).
\begin{equation}\label{eq:second-order-split}
\operatorname{E}_{A+B}(h)=\operatorname{E}_{A}(h/2)\operatorname{E}_{B}(h)\operatorname{E}_{A}(h/2)+\operatorname{O}\left(h^{2}\right)
\end{equation}
The sub-Hamiltonian $A$ is evolved in two steps of $h/2$ and the sub-Hamiltonian $B$ is evolved in a single step $h$. We can take advantage of this property of the splitting formula by dividing terms associated with fast interactions into $A$ and terms associated with slow interactions into $B$. We can proceed by applying this splitting procedure to different sub-Hamiltonians multiple times, thereby constructing an integrator that evaluates parts of the Hamiltonian at $h$, $h/2$, $h/4$ etc, similarly to the power of two hierarchy used in block time step schemes.  This approach was followed in \cite{Pelupessy:2012if}, below we will introduce some notation and give a rough derivation of the integrators there.

Hamiltonians consisting of a single momentum term and Hamiltonians consisting of a single potential term have analytic solutions. For a momentum term of the $i$-th particle
\begin{equation}
H_{i}(\mathbf{p}_{i},\mathbf{q}_{i})=T_{i}=\frac{\left\langle \mathbf{p}_{i},\mathbf{p}_{i}\right\rangle }{2m_{i}}
\end{equation}
the solution consists of updating the position of the $i$-th particle under the assumption of constant velocity for a time period of $h$ (all positions except the position of the $i$-th particle and the momenta of all particles remain unchanged). 
\begin{equation}
\label{eq:drift}
\mathbf{q}_{i}(t+h)=\mathbf{q}_{i}(t)+h\mathbf{v}_{i}(t)
\end{equation}
We call the time evolution operator for the momentum term of the $i$-th particle the \emph{drift operator} and write $D_{h,T_{i}}$.

For a single potential term between particles $i$ and $j$
\begin{equation}
\label{eq:kick}
H_{ij}(\mathbf{p}_{i},\mathbf{q}_{i})=V_{ij}=-G\frac{m_{i}m_{j}}{\sqrt{q_{ij}^{2}+\varepsilon^{2}}}
\end{equation}
the solution consists of updating the momenta of the $i$-th and $j$-th particles under the assumption of constant force for a time period of $h$ (all momenta except the momenta of the $i$-th and $j$-th particles and the positions of all particles remain unchanged). 
\begin{eqnarray}
\label{eq:kickij}
\mathbf{p}_{i}(t+h) & = & \mathbf{p}_{i}(t)+h\mathbf{F}_{ij}(t)\\
\label{eq:kickji}
\mathbf{p}_{j}(t+h) & = & \mathbf{p}_{j}(t)+h\mathbf{F}_{ji}(t)
\end{eqnarray}
We call the time evolution operator for the potential term between the $i$-th and $j$-th particles the kick operator and write $K_{h,V_{ij}}$.

In addition to the kick and drift operators, the two-body Hamiltonian
\begin{equation}
\label{eq:kepler}
H_{ij}(\mathbf{p}_{i},\mathbf{q}_{i})=T_{i}+T_{j}+V_{ij}=
 {\frac{\langle \mathbf{p}_{i},\mathbf{p}_{i}\rangle }{2m_{i}}} + 
 {\frac{\langle \mathbf{p}_{j},\mathbf{p}_{j}\rangle }{2m_{j}}} -
 G\frac{m_{i}m_{j}}{\sqrt{q_{ij}^{2}+\varepsilon^{2}}}
\end{equation}
is solved (semi-) analytically by the Kepler solution\footnote{even the case with $\varepsilon \ne 0$ can be solved in a universal variable formulation (Ferrari, priv. comm.)}. 

In \cite{Pelupessy:2012if}, we derive multiple integrators that recursively and adaptively split the system Hamiltonian through the second-order splitting formula \eqref{eq:second-order-split}. At every step in the recursion, all particles under consideration are divided into a slow set $S$ and a fast set $F$ by comparing the particle-specific time step function $\tau(i)$ to a pivot time step $h$.
\begin{eqnarray}
S & = & \left\{ i\in1\ldots N:\,\tau(i)\geq h\right\} \\
F & = & \left\{ i\in1\ldots N:\,\tau(i)<h\right\} 
\end{eqnarray}
Using the two sets $S$ and $F$, we can rewrite the system Hamiltonian as follows.
\begin{eqnarray}
H & = & H_{S}+H_{F}+V_{SF}\label{eq:SF-splitting-rule}
\end{eqnarray}
The sub-Hamiltonian $H_{S}$ can be thought of as a ``closed Hamiltonian'' of the particles in $S$. Specifically, it consists of all drifts of particles in $S$ and all kicks where both participating particles are in $S$. The same property holds for the sub-Hamiltonian $H_{F}$ and the particles in $F$. The mixed term $V_{SF}$ contains all kicks where one particle is in $S$ and the other is in $F$.

We proceed by applying the second-order splitting rule \eqref{eq:second-order-split}:
\begin{eqnarray}
\operatorname{E}_{h,H} & = & \operatorname{E}_{h,H_{S}+H_{F}+V_{SF}}\\
 & \approx & \operatorname{E}_{h/2,H_{F}}\,\operatorname{E}_{h,H_{S}+V_{SF}}\operatorname{E}_{h/2,H_{F}}\label{eq:HOLD}
\end{eqnarray}
(this is not the only conceivable approximation).
The sub-Hamiltonian $H_{F}$ is closed, and consists of particles where $\tau(i)<h$. We integrate $H_{F}$ by recursively applying the entire ``slow/fast'' partitioning, but using a smaller pivot $h/2$. In contrast, both $H_{S}=T_{S}+V_{S}$ and $V_{SF}$ are explicitly decomposed into individual kicks and drifts which are applied using the current pivot time step $h$. We refer to this particular choice as the HOLD method (since it 'holds' $V_{SF}$ for evaluation at the slow timestep).
\begin{eqnarray}
\operatorname{E}_{h,H_{S}+V_{SF}} & = & \operatorname{E}_{h,T_{S}+V_{S}+V_{SF}}\\
 & \approx & \operatorname{D}_{h/2,T_{S}}\operatorname{K}_{h,V_{S}+V_{SF}}\operatorname{D}_{h/2,T_{S}}
\end{eqnarray}
The pivot time step $h$ is halved with each consecutive partitioning, and the recursion terminates when all remaining particles are placed into the $S$ set.

As noted previously, recursively and adaptively splitting the system Hamiltonian using the second order splitting rule (Eq \ref{eq:second-order-split}) is similar to conventional block time steps. Both approaches evolve different parts of the system using time steps that belong to a power of two hierarchy. However, the Hamiltonian splitting method derived above evaluates pairwise particle forces symmetrically in the sense that a ``kick'' from particle $i$ to particle $j$ (Eq \ref{eq:kickij}) is always paired with an opposite kick from particle $j$ to particle $i$ (Eq \ref{eq:kickji}). Furthermore, the kicks acting upon a particle at any given timestep typically correspond to partial forces only. This is in contrast to conventional block time steps where we always calculate the {\em total} force acting on a particle at the frequency determined by the particle-specific time step criteria $\tau(i)$
\begin{eqnarray}
\label{eq:Hamiltonian}
\mathbf{p}_{i}(t+h) & = & \mathbf{p}_{i}(t)+h\underset{j=1, j\neq i}{\overset{N}{\sum}}\mathbf{F}^{\star}_{ij}(t)
\end{eqnarray}
where, $\mathbf{F}^{\star}_{ij}(t)$ is the force acting on particle $i$ due to particle $j$, derived from extrapolated positions if necessary. Specifically, in situations where the position of particle $j$ has not been calculated for time $t$, we calculate the force by extrapolating the position at $t$ from the last known position. This can happen when particle $i$ is assigned a smaller time step than particle $j$. We refer to this method as BLOCK, and include it as a reference in our numerical tests to determine whether more ``aggressive'' splitting methods (such as HOLD) reduce the number of kicks and drifts while maintaining the accuracy of the solution.

The HOLD method evolves all kicks between fast particles at the fast time step. This is inefficient in the presence of isolated fast subsystems, as interactions between particles that belong to different subsystems could be evolved at a slower time step. As an extreme example, consider a Plummer sphere with each star being replaced by a stable hard binary. Here, every star has a close binary interaction that needs to be evaluated at a fast time step. However, the HOLD integrator will in this case integrate all interactions, including long-range interactions between stars in different binaries at a time step determined the binary interactions. The behaviour of the method becomes equivalent to evolving the entire system with a shared global time step!

In addition to the dramatic example just discussed, the same inefficiency --- evaluating long-range interactions between isolated fast subsystems at time steps determined by fast interactions inside the subsystems --- can manifest itself in other situations, such as the following.
\begin{itemize}
\item In a system with multiple globular clusters, each individual globular cluster is a subsystem.
\item In a globular cluster with planets around some of the stars, each star with planets is a subsystem.
\item In a single globular cluster, each close encounter between two or more stars is a subsystem.
\end{itemize}

\subsection{Hamiltonian splitting with connected components}\label{sec:cc-split}

The partitioning used in the HOLD method is based on a \emph{particle-specific} time step criteria $\tau(i)$, which by definition cannot separate slow and fast interactions in situations where all particle-specific time steps have the same (fast) value. We therefore introduce the \emph{interaction-specific} time step criterion $\tau(i,j)$
\begin{equation}
\label{eq:tauij}
\tau(i,j) = \eta \min \left(
	\frac{\tau_\textrm{freefall}(i,j)}{\left(1-\frac{1}{2}\frac{d\tau_\textrm{freefall}(i,j)}{dt}\right)},
	\frac{\tau_\textrm{flyby}(i,j)}{\left(1-\frac{1}{2}\frac{d\tau_\textrm{flyby}(i,j)}{dt}\right)}
\right)
\end{equation}
where $\tau_\textrm{freefall}(i,j)$ and $\tau_\textrm{flyby}(i,j)$ are proportional to the interparticle free-fall and interparticle flyby times as defined by eqs (13) and (16) in \cite{Pelupessy:2012if}, and $\eta$ is an accuracy parameter.

We split the system Hamiltonian using the connected components\cite[Sec B.4]{Cormen:2001uw} of the undirected graph generated by the time step criteria $\tau(i,j)$. Specifically, the particles of the system correspond to the vertices of the graph, and there is a edge between particles $i$ and $j$ if their interaction cannot be evaluated at the threshold time step $h$.
\begin{equation}
\tau(i,j)<h
\end{equation}
\begin{figure*}
\includegraphics[width=0.32\textwidth]{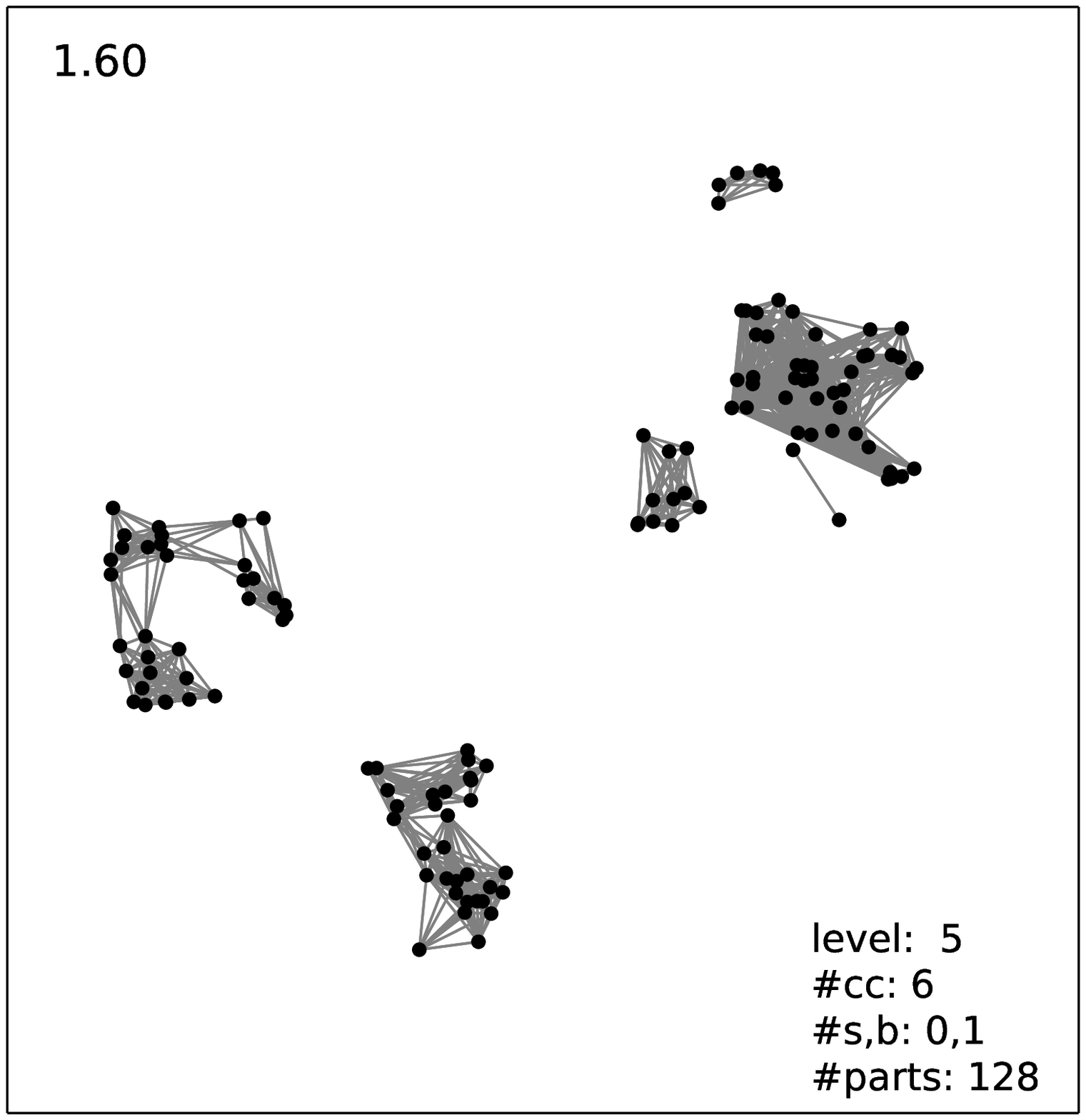}
\includegraphics[width=0.32\textwidth]{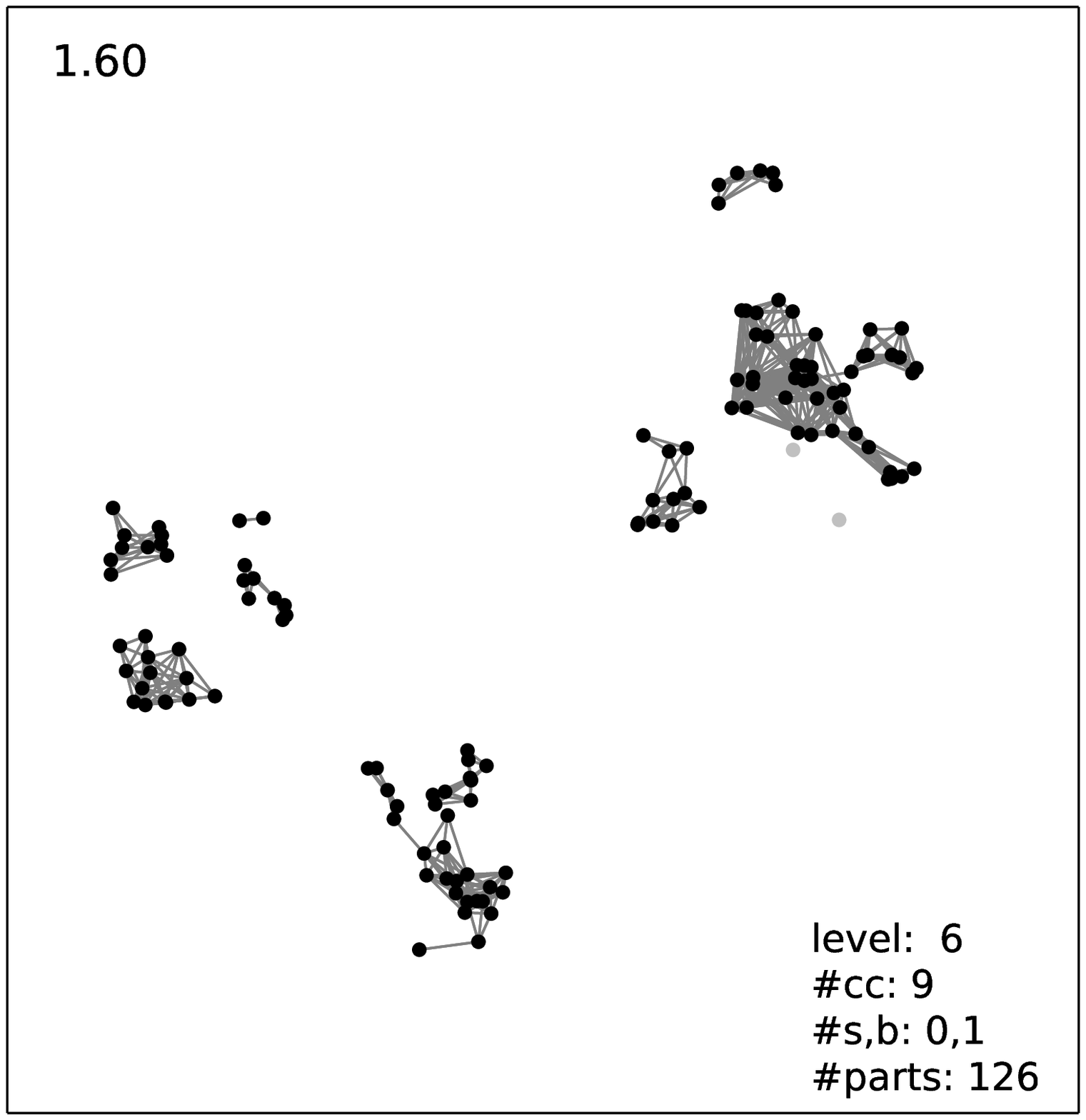}
\includegraphics[width=0.32\textwidth]{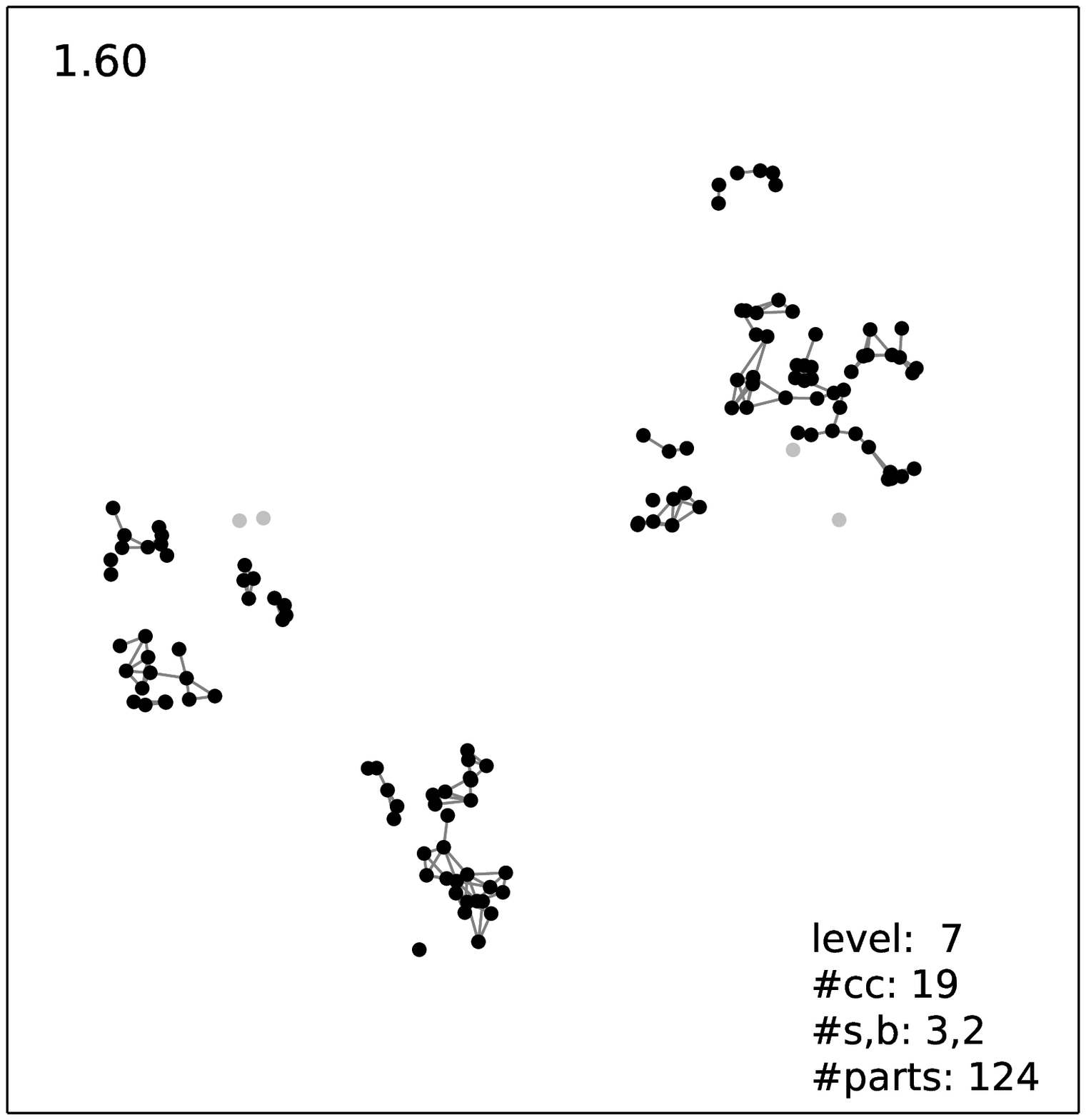}

\includegraphics[width=0.32\textwidth]{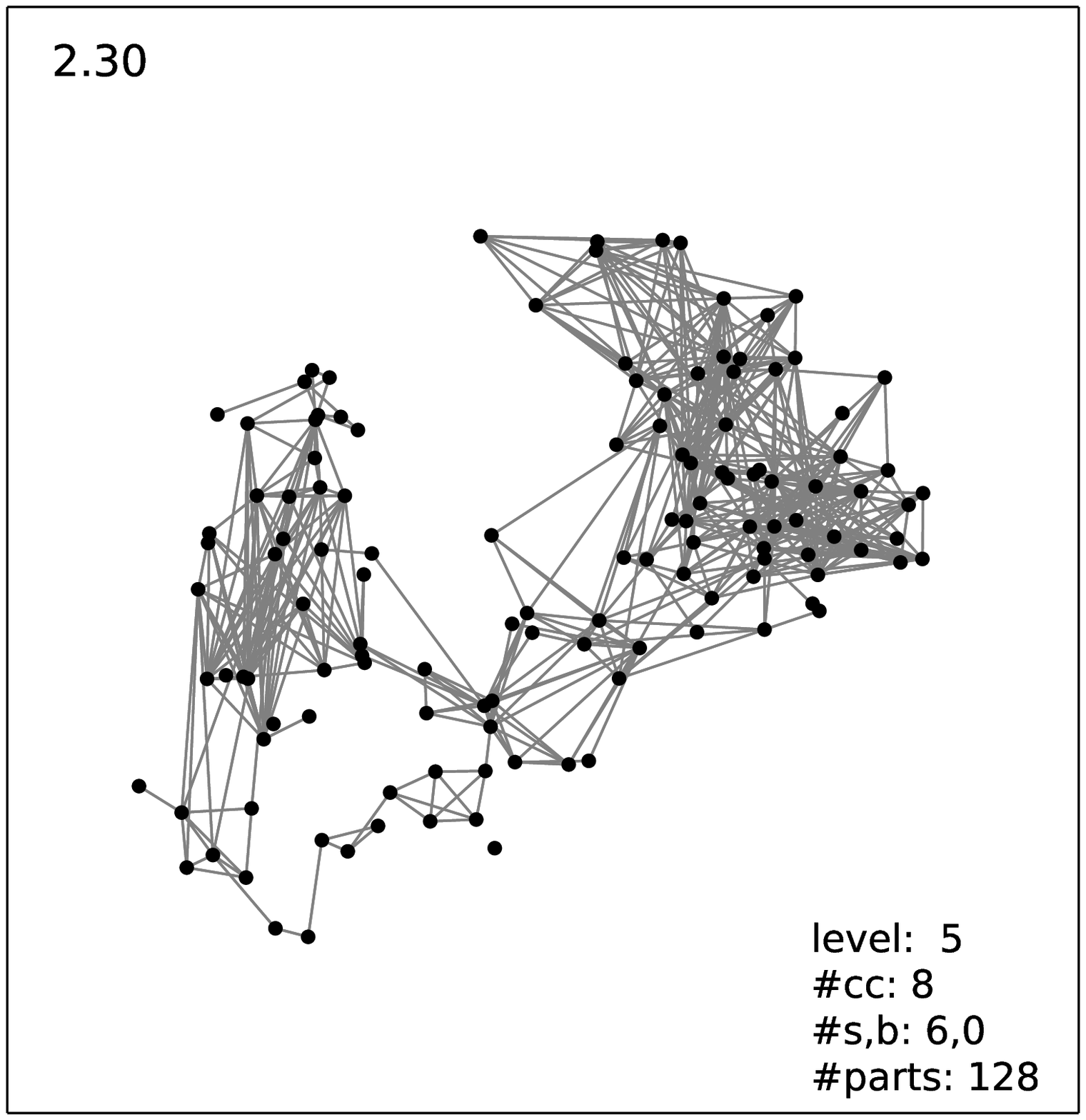}
\includegraphics[width=0.32\textwidth]{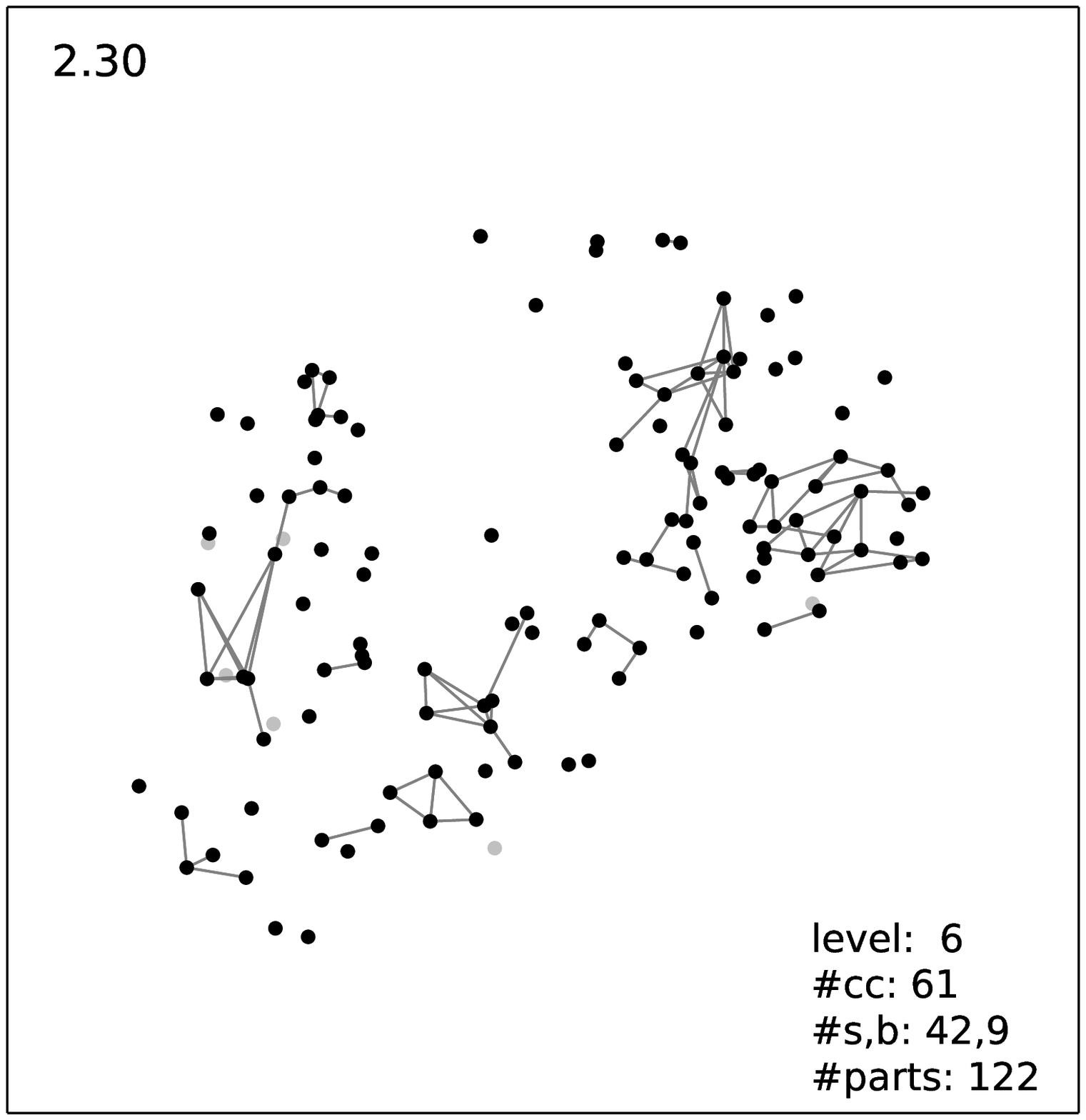}
\includegraphics[width=0.32\textwidth]{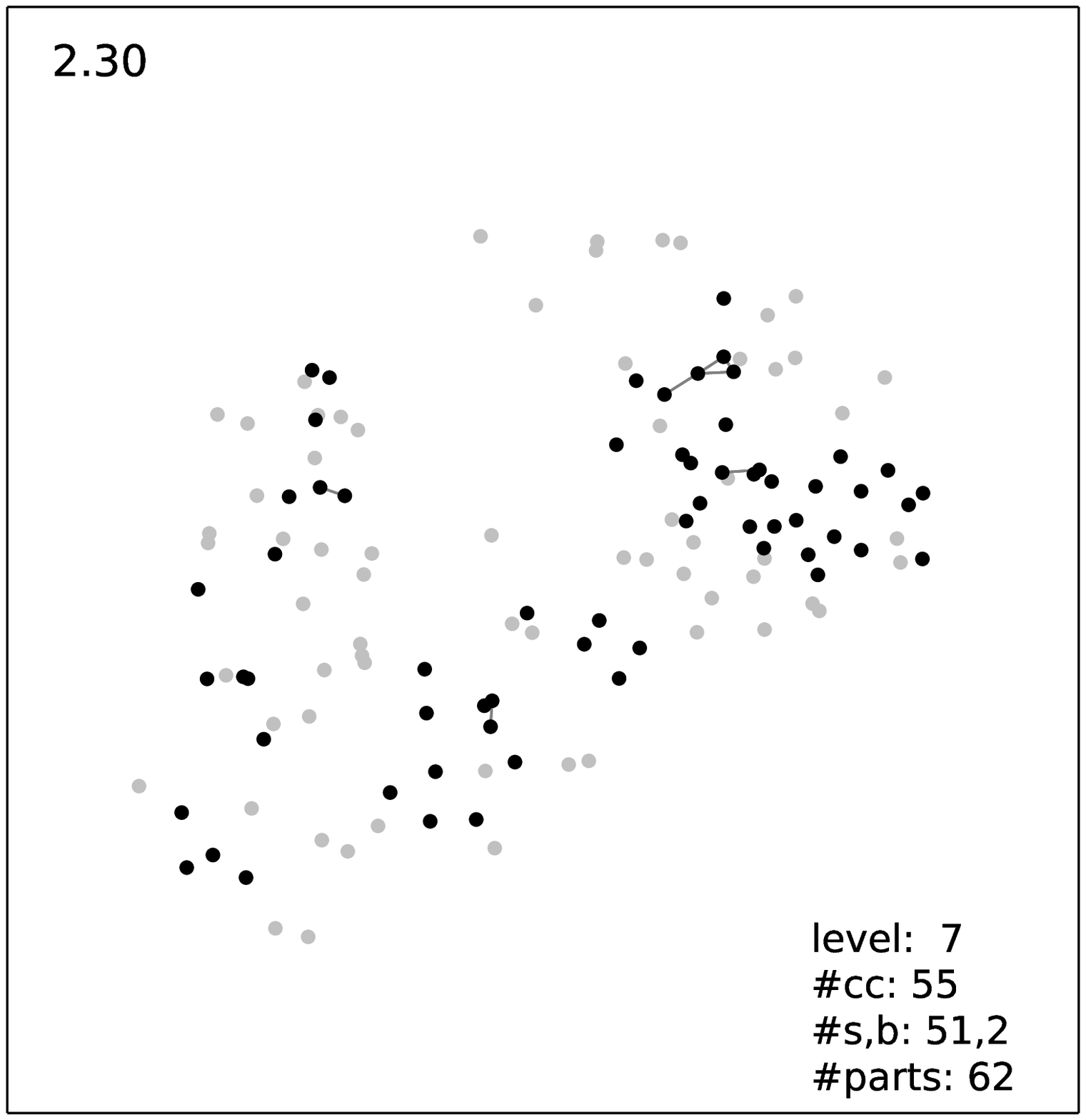}

\includegraphics[width=0.32\textwidth]{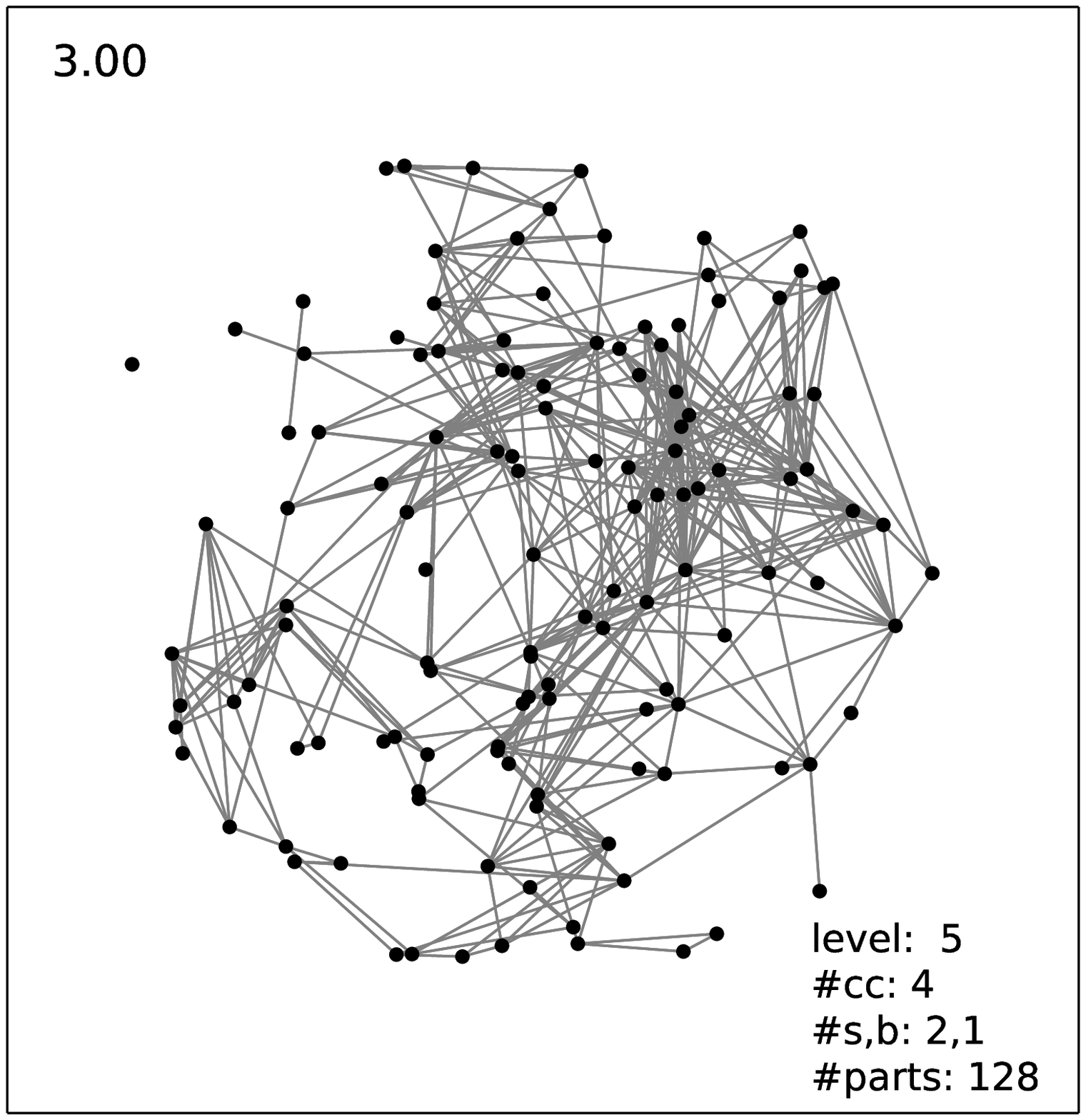}
\includegraphics[width=0.32\textwidth]{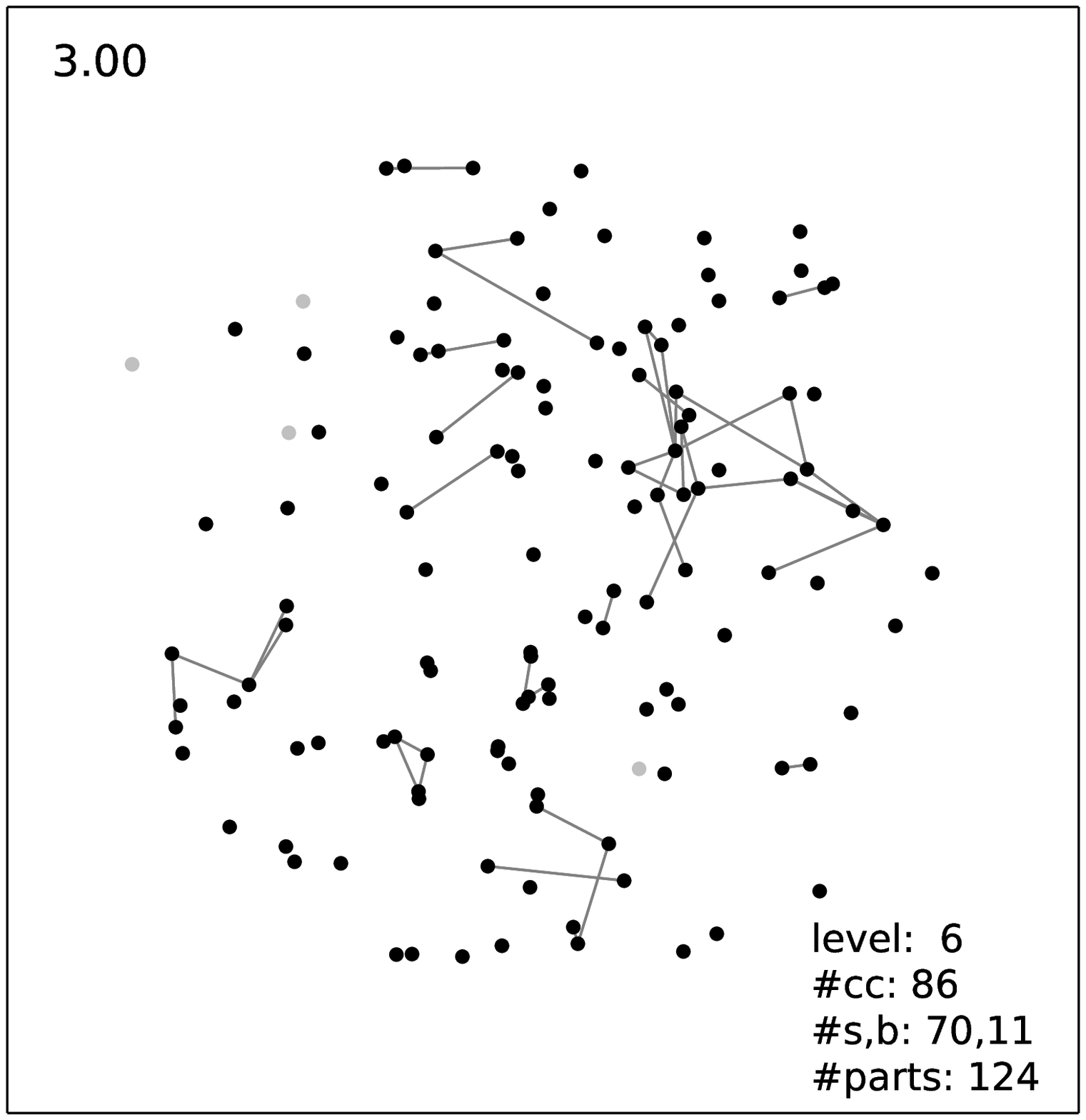}
\includegraphics[width=0.32\textwidth]{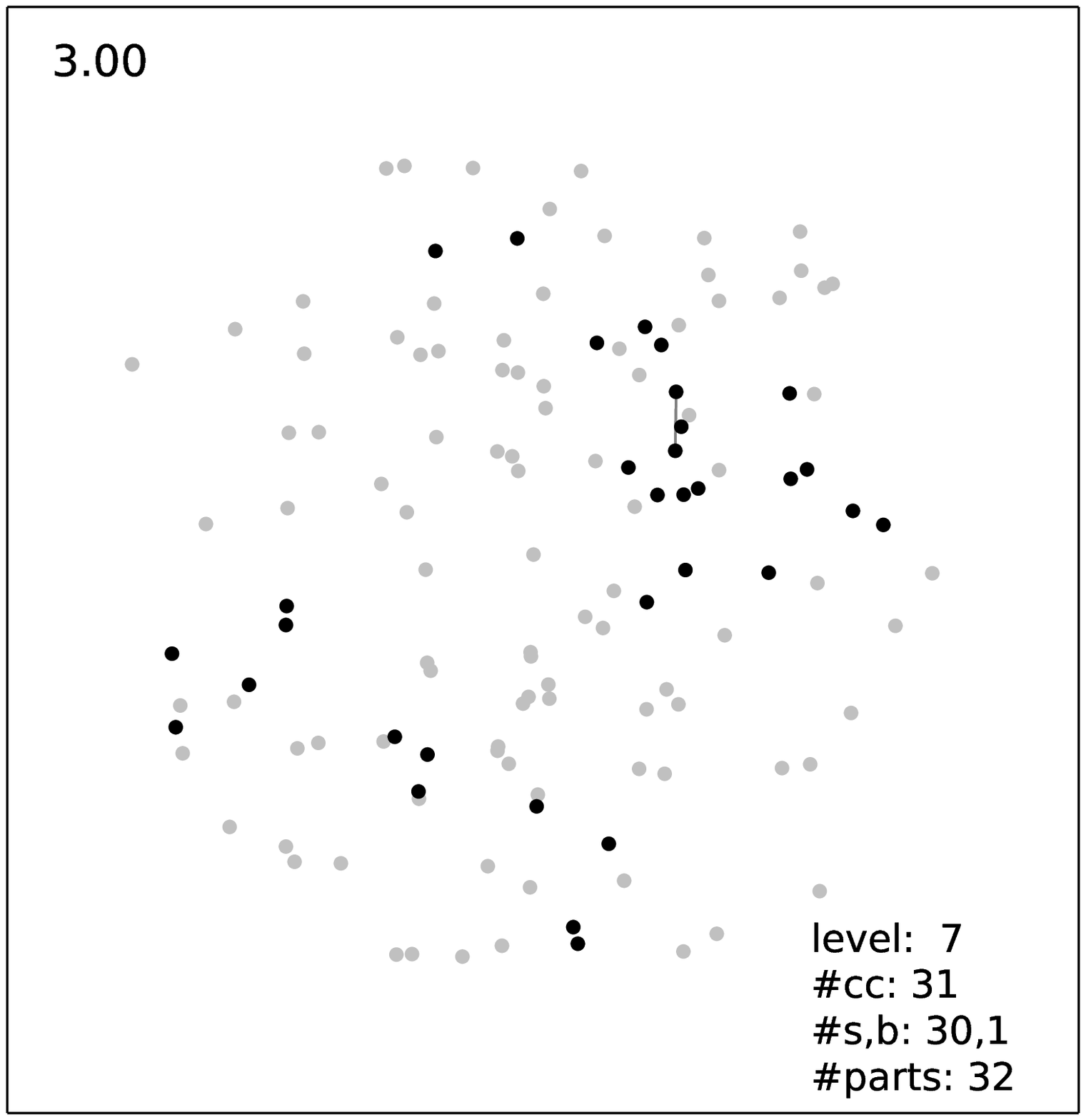}

\caption{\label{fig:time-step-graph} Time step graphs generated by $\tau(i,j)$ at different levels of the time step hierarchy (left to right) for three values (top to bottom rows) of the fractal dimension. We plot particles that have been passed on as a part of a connected component from the previous time step level as black, grey points indicate points that are inactive on a given level (because they formed a single or binary component at a lower level). Thin grey lines indicate interactions with $\tau(i,j)<h$. Indicated in each frame are the fractal dimension (top left), and (on the bottom right) the level in the hierarchy, the number of connected components (cc) at this level, as well as the number of components that are single (s) and binary (b), and the total number of particles. }
\end{figure*}
Figure \ref{fig:time-step-graph} we visualises the time step graphs at varying values of the pivot time step $h$ for three different fractal initial conditions with different fractal dimension (described further in Section \ref{sec:tests}). As the pivot time step $h$ decreases, the set of interactions (and associated particles) that \emph{cannot} be evaluated at the current pivot time step gradually decreases as well. Although for visualisation purposes, we plot the time step graph of the entire system for varying $h$, the CC (\emph{Connected Components}) splitting method we are about to introduce typically calculates connected components for the entire system only once, at the largest pivot time step. At smaller pivot time steps, the connected components search is only calculated for parts of the system. The intuition behind this partitioning comes from clustering by maximising the margin between individual clusters as described in \cite{Duan:2009kh}.

Given a fixed pivot time step $h$, let the sets $C_{i},\, i=1\ldots K$ contain vertices of $K$ non-trivial connected components, and the set $R$ (``remainder set'') contain all particles in trivial connected components.\footnote{A \emph{trivial connected component} is a connected component with exactly one vertex and a \emph{non-trivial connected component} is a connected component with at least two vertices.} Based on the particle sets $C_{i}$ and $R$, we rewrite the Hamiltonian of the system in the following form
\begin{equation}
H = H_{C}+H_{R}+V_{CC}+V_{CR}
\end{equation}
where the individual terms are defined as follows.
\begin{eqnarray}
H_{C} & = & \underset{i=1}{\overset{K}{\sum}}H_{C_{i}}=\underset{i=1}{\overset{K}{\sum}}\left(\underset{j\in C_{i}}{\overset{}{\sum}}T_{j}+\underset{\substack{j,k\in C_{i}\\
i<j
}
}{\overset{}{\sum}}V_{jk}\right)\\
H_{R} & = & T_{R}+V_{R}=\underset{i\in R}{\overset{}{\sum}}T_{i}+\underset{\substack{i,j\in R\\
i<j
}
}{\overset{}{\sum}}V_{ij}\\
V_{CC} & = & \underset{\substack{i,j=1\\
i<j
}
}{\overset{K}{\sum}}V_{C_{i}C_{j}}=\underset{\substack{i,j=1\\
i<j
}
}{\overset{K}{\sum}}\left(\underset{\substack{k\in C_{i}\\
l\in C_{j}
}
}{\overset{}{\sum}}V_{kl}\right)\\
V_{CR} & = & \underset{i=1}{\overset{K}{\sum}}V_{C_{i}R}=\underset{i=1}{\overset{K}{\sum}}\left(\underset{\substack{j\in C_{i}\\
k\in R
}
}{\overset{}{\sum}}V_{jk}\right)
\end{eqnarray}
The term $H_{C}$ is the sum of all closed Hamiltonians $H_{C_{i}}$, each corresponding to one of the $K$ connected components. In every $H_{C_{i}}$ all drifts and some kicks cannot be evolved at the time step $h$ without violating the time step criteria. The term $H_{R}$ consists of the closed Hamiltonian formed by all of the particles in the rest system. All drifts and kicks in $H_R$ can be evolved at the current time step $h$. 

The term $V_{CC}$ contains all kicks between particles that are in \emph{different} connected components. These kicks can be evaluated at the time step $h$. We explicitly point out that $V_{CC}$ explicitly contains the terms that are evolved inefficiently in the HOLD method. Similarly, $V_{CR}$ contains all kicks where one of the particles is in a connected component $C_i$, and the other is in the rest set $R$.

We split the system Hamiltonian $H$ by applying the second-order splitting rule \eqref{eq:second-order-split}:
\begin{eqnarray}
\operatorname{E}_{h,H} & = & \operatorname{E}_{h,H_{C}+H_{R}+V_{CC}+V_{CR}}\\
 & \approx & \operatorname{E}_{h/2,H_{C}}\,\operatorname{E}_{h,H_{R}+V_{CC}+V_{CR}}\operatorname{E}_{h/2,H_{C}}
\end{eqnarray}
such that individual connected components $C_i$ are independently evolved at a higher pivot time step $h/2$ via recursion.
\begin{equation}
\operatorname{E}_{h/2,H_{C}}=\overset{K}{\underset{i=1}{\prod}}\operatorname{E}_{h/2,H_{C_{i}}}
\end{equation}
All remaining terms (including $V_{CC}$) are decomposed into individual drifts and kicks using the second-order splitting rule \eqref{eq:second-order-split}.
\begin{eqnarray}
\operatorname{E}_{h,H_{R}+V_{CC}+V_{CR}} & = & \operatorname{E}_{h,T_{R}+V_{R}+V_{CC}+V_{CR}}\\
 & \approx & \operatorname{D}_{h/2,T_{R}}\operatorname{K}_{h,V_{R}+V_{CC}+V_{CR}}\operatorname{D}_{h/2,T_{R}}\\
 & = & \operatorname{D}_{h/2,T_{R}}\operatorname{K}_{h,V_{R}}\operatorname{K}_{h,V_{CC}}\operatorname{K}_{h,V_{CR}}\operatorname{D}_{h/2,T_{R}}
\end{eqnarray}
As with the HOLD method, the pivot time step $h$ is halved at each successive partitioning such that at some point, all remaining particles are in the remainder set $R$.

Finally, the partitioning can lead to situations where $H_{C_{i}}$ is a two-body Hamiltonian (Eq.~\ref{eq:kepler}). We can use this property by integrating these cases with a dedicated a Kepler solver (we discuss this further in Section \ref{sec:plum}).

\subsection{Implementation}

\begin{figure*}
\begin{verbatim}
evolve_cc(H, h):
    // split_cc() decomposes particles in H (eq 25) into:
    // 1) K non-trivial connected components C_1..C_K
    // 2) Rest set R 
    (C_1..C_K, R) = split_cc(H, h);

    // Independently integrate every C_i at reduced pivot time step h/2 (eq 27)
    for C_i in C_1..C_K:
        evolve_cc(C_i, h/2)

    // Apply drifts and kicks at current pivot time step h (eq 30)
    drift(R, h/2) // evolves T_R
    kick(R, R, h) // evolves V_RR
    kick(C_1..C_K, C_1..C_K, h) // evolves V_CC (eq 23)
    kick(C_1..C_K, R, h) // evolves V_CR (eq 24)
    drift(R, h/2) // evolves T_R

    // Independently integrate every C_i at reduced pivot time step h/2 (eq 27)
    for C_i in C_1..C_K:
        evolve_cc(C_i, h/2)
\end{verbatim} 
\caption{\label{fig:cc-pseudo-code}Pseudocode for a second-order CC splitting routine for integrating a set of particles {\tt H} for time step {\tt h}.
}
\end{figure*}

We implemented the CC split in the HUAYNO code, which is freely available as a part of the AMUSE framework. Figure \ref{fig:cc-pseudo-code} sketches the main routine of the CC integrator in pseudocode, including explicit references to the corresponding equations and variables used in the derivation of the method (Section \ref{sec:cc-split}).

Subroutines and data structures that store the system state, calculate time steps, apply kicks and drifts to groups of particles, and gather statistics, are shared with other integrators such as the HOLD method. All particle states are kept in a contiguous block of memory. The connected component algorithm is implemented as a breadth-first search. It reshuffles particle states such that particles in the same connected component or rest set are kept adjacent to each other. Connected components are represented by a start and an end pointer to the contiguous array of particle states.

The time complexity of the connected component decomposition for $N$ particles has an upper bound of $O(N^2)$. This matches the time complexity of the splitting step of the HOLD method --- while the actual shuffling of the particles into $S$ and $F$ sets is $O(N)$, this division is based on the preceding step of calculating particle-based time steps $\tau(i)$ for all particles, which is $O(N^2)$.

For the special case where all interactions between the $N$ particles are below the threshold $h$, the complexity of the connected components decomposition is $O(N)$. This can happen multiple times (at consecutive recursion levels) when the initial value of the pivot time step $h$ is sufficiently large. Figuratively, if particle $X$ has a known connected component while particle $Y$ is unassigned, we can assign particle $Y$ to the connected component of particle $X$ based on a single time step evaluation $\tau(X,Y)$. A key step of the connected components search is choosing a particle $X$ with a known connected component, followed by assigning the membership of $X$ to all unassigned particles $U_k$ where $\tau(X, U_k)<h$. For the special case under consideration, a single iteration of this step is sufficient to assign membership to all particles (irrespective of the choice of the initial particle $X$), leading to a time complexity of $O(N)$. Further, while the splitting step is bounded from above by $O(N^2)$ for both HOLD and CC, the HOLD split always calculates time steps for all interactions. This is not the case with the CC method, and numerical tests in Section \ref{sec:tests} indicate that the reduction in time step evaluations does translate into improved performance.

\section{Tests}\label{sec:tests}

\begin{table*}
\begin{tabular}{|p{0.2\linewidth}|p{0.8\linewidth}|}
\hline 
Method & Description\tabularnewline
\hline 
\hline 
BLOCK & Conventional particle-based block time steps --- positions of particles in lower time step bins are extrapolated when calculating the movement of particles in faster time step bins. \tabularnewline
\hline 
HOLD &  Individual timestepping method based on Hamiltonian splitting. Particles in different timestep bins interact by exchanging symmetric kicks 
\citep{Pelupessy:2012if}. \tabularnewline
\hline 
CC & An implementation of the connected components splitting (\mbox{Section \ref{sec:cc-split}}). Iterative partitioning based on the connected 
components of the graph generated by the pairwise timestep criterion. \tabularnewline
\hline 
CC\_KEPLER & An extension of the CC method that uses a Kepler solver to evolve connected components with two particles (Section \ref{sec:plum}).\tabularnewline
\hline 
\end{tabular}

\caption{\label{table:methods}
An overview of the integrators used in the numerical tests. The implementations share a significant amount of the code and data structures used for storing system state, calculating time steps, and evaluating (partial) forces.
}
\end{table*}

We present results of numerical experiments of the connected components (CC) splitting method described in the previous section.
We confirm that the CC method works as intended conceptually by comparing to alternative Hamiltonian splitting methods (see Table \ref{table:methods} for an overview). Specifically, we demonstrate that the connected components search does not use excessive computational resources, reduces the number of elementary operations (kick, drift and time step evaluations) while maintaining the accuracy of the solution, and performs particularly well on multi-scale problems. Finally, we compare the CC method to established $N$-body codes.
We use $N$-body units as described in \cite{1986LNP...267..233H}.

\subsection{Smoothed Plummer sphere test}\label{sec:smoothplum}

\begin{figure*}
\centering
\includegraphics[width=0.8\textwidth]{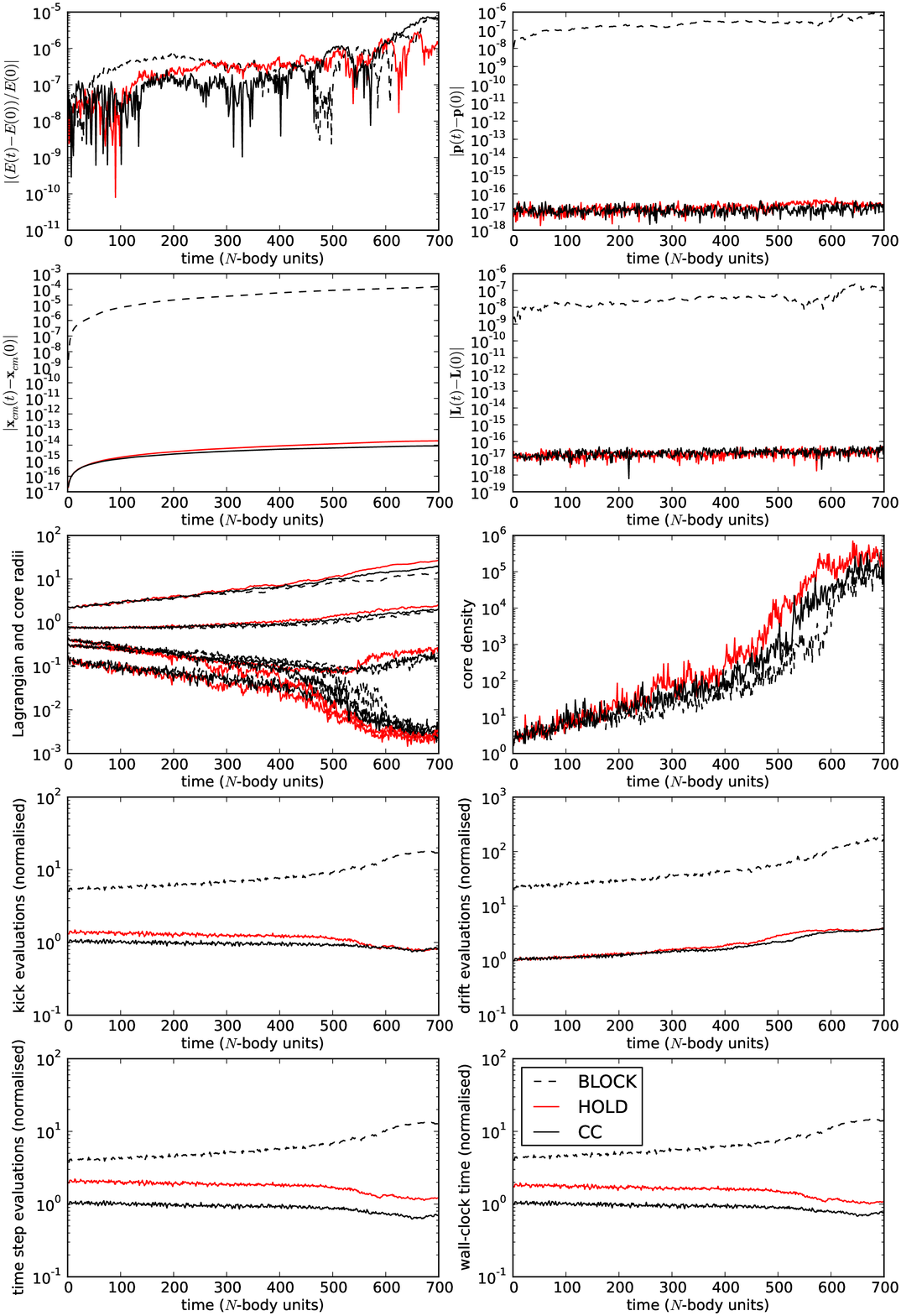}
\caption{\label{fig:plummer-unsoftened} A comparison of the BLOCK, HOLD and CC methods on integrating an softened $1024$-body Plummer sphere: conservation of the integrals of motion (top two rows), evolution of the mass distribution of the solution (middle row) and performance metrics (two bottom rows). Lagrangian radii are plotted for 90\%, 50\%,  10\% and 1\% of the system mass. The performance metrics are normalised to the CC method at the start of the simulation. The CC method performs $1.1 \times 10^9$ kick, $6.6 \times 10^8$ time step, and $5.0 \times 10^6$ drift evaluations, taking 62 seconds for the first global time step ($1/512$-th of the simulation)
 on a laptop with a 1.3GHz Intel Core i5 processor).
}
\end{figure*}

We begin by integrating an equal-mass $1024$-body Plummer sphere with softening ($\varepsilon=1/256$) for $700$ $N$-body time units using a time step accuracy parameter of $\eta=0.01$. We choose initial velocities such that the Plummer sphere is in a dynamic equilibrium. This setup is chosen to match the long-term integration tests in \cite[their section 3.2]{Nitadori:2008gt}.

Figure \ref{fig:plummer-unsoftened} visualises the conservation of the integrals of motion, the time evolution of the mass distribution, and performance metrics. While all three methods show similar energy conservation properties, only HOLD and CC maintain centre of mass, linear momentum and angular momentum near machine precision. As noted previously in \cite{Pelupessy:2012if}, this is caused by unsynchronised kicks which are only present in the BLOCK scheme. The solutions obtained by all three methods reproduce known results in terms of Lagrangian radii, the core radius and the core density. The CC scheme is about twice as fast than the HOLD scheme at the beginning of the simulation, and remains the fastest scheme throughout the run. The overall runtime measurements correlate with the number of time step formula evaluations and, to a lesser extent, the number of kick and drift formula evaluations. This indicates that the improved runtime is attributable to a reduction of time step, kick and drift formula evaluations. 

\begin{figure*}
\centering
\includegraphics[width=0.85\textwidth]{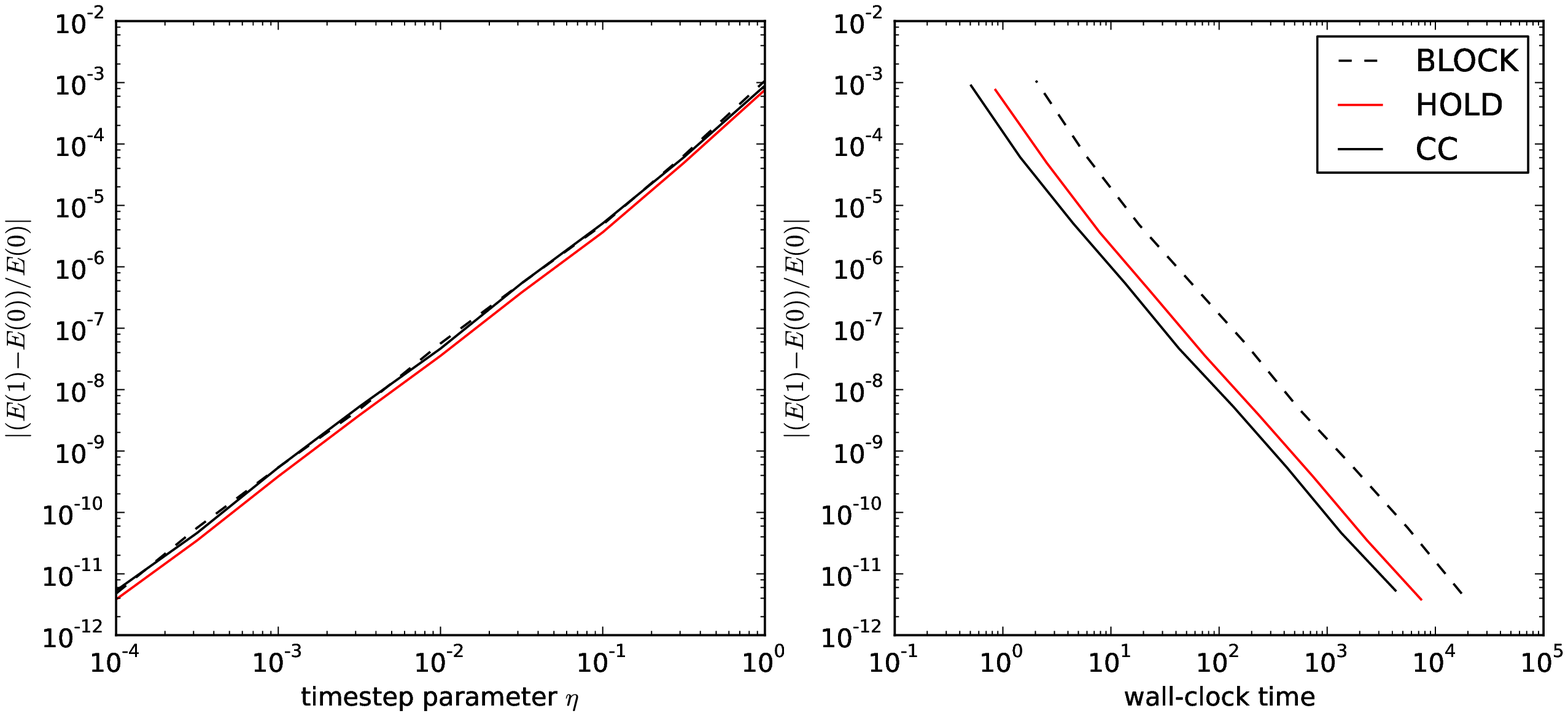}
\caption{\label{fig:second-order}
Left: time step accuracy parameter $\eta$ vs energy error for integrating a $1024$-body Plummer sphere for $1$ $N$-body units. Right: corresponding wall-clock time vs energy error from the same set of tests.
}
\end{figure*}

The left plot of Figure \ref{fig:second-order} visualises energy error of evolving the softened Plummer sphere as described previously, but for $1$ $N$-body units and under varying time step accuracy $\eta$. As predicted, all three methods show second order behaviour. On the corresponding wall-clock time vs energy error plot on the right CC consistently outperforms HOLD, followed by BLOCK. We emphasise that the Plummer sphere is a spherically symmetric configuration with a smoothly changing mass distribution, and a non-zero softening length $\varepsilon$ sets an upper limit on the hardness of the binaries that can form during the simulation. Hence, we would not expect the CC scheme to have a significant advantage over the HOLD method.	

\begin{figure*}
\centering
\includegraphics[width=0.8\textwidth]{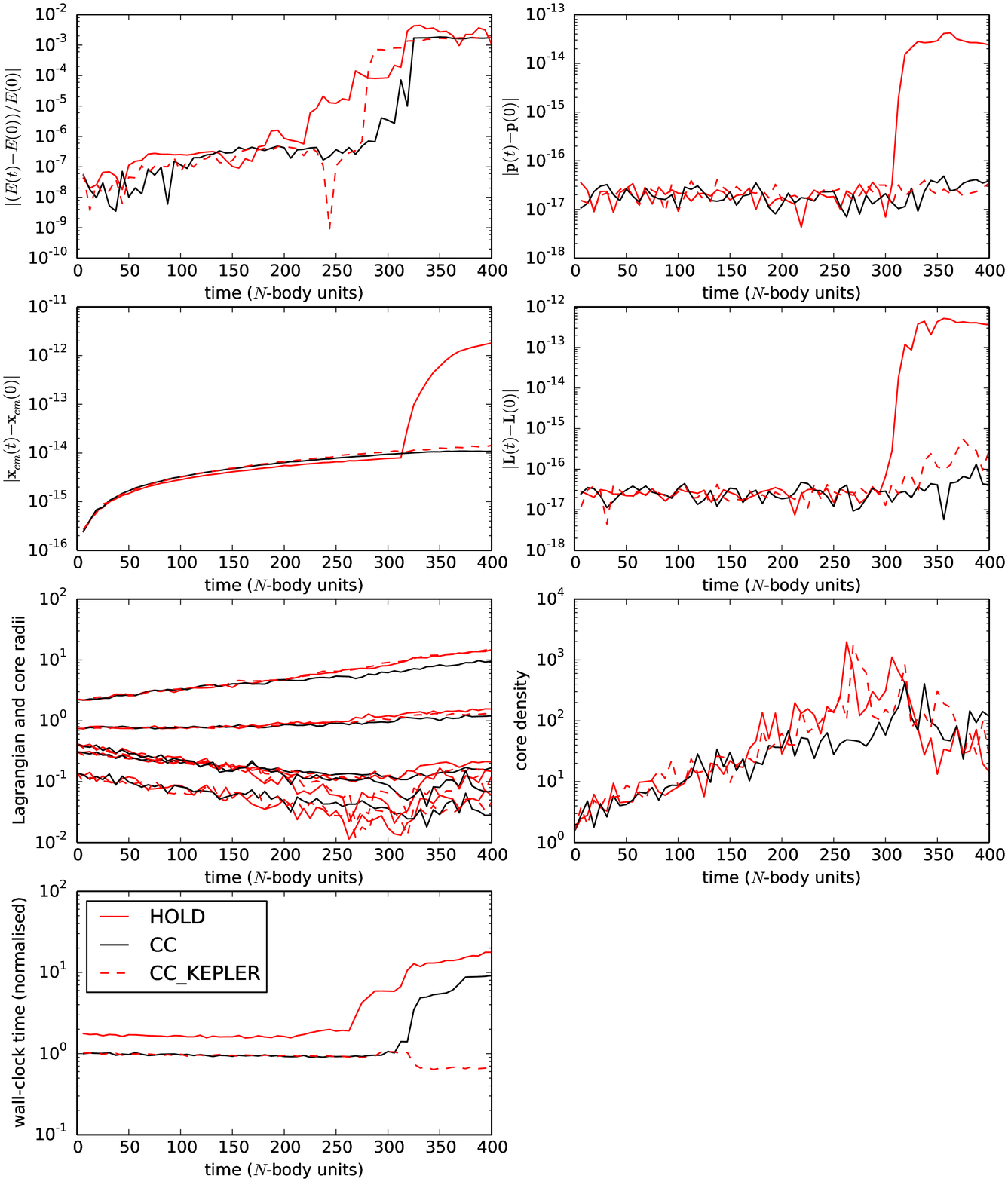}
\caption{\label{fig:plummer-core-collapse} A comparison of the HOLD, CC and CC\_KEPLER methods on integrating an unsoftened $1024$-body Plummer sphere through core collapse: conservation of the integrals of motion (top two rows), evolution of the mass distribution of the solution (third row) and wall-clock time (bottom left). Lagrangian radii are plotted for 90\%, 50\%,  10\% and 1\% of the system mass. Wall-clock times are normalised by CC\_KEPLER at the first global time step ($1/64$-th of the simulation). On a laptop with a 1.3GHz Intel Core i5 processor, CC\_KEPLER integrates the first global time step in roughly $5$ minutes while the entire simulation takes about $7$ hours.
}
\end{figure*}

\begin{figure}
\centering
\includegraphics[width=0.49\textwidth]{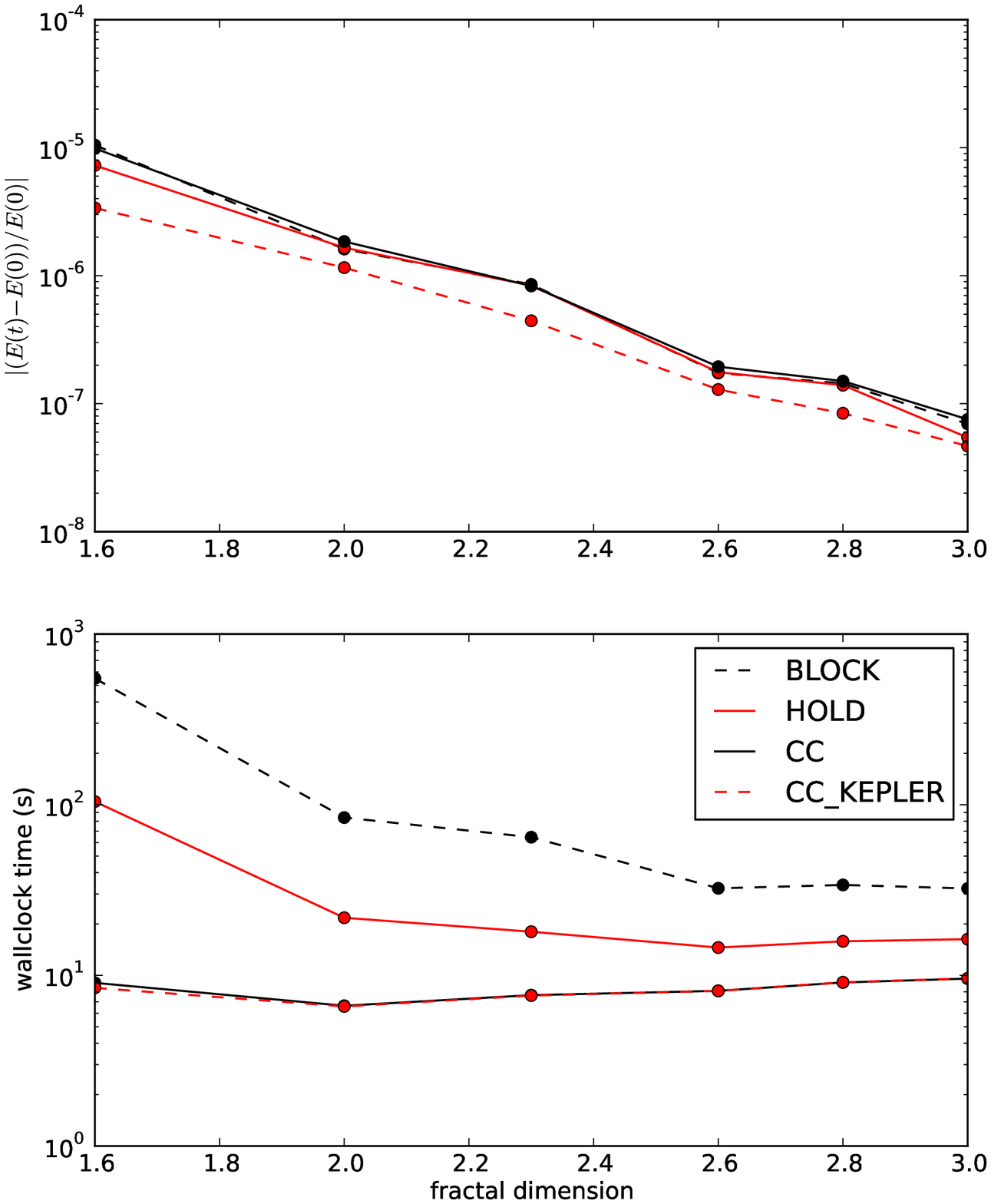}
\caption[]{\label{fig:fractal-initial-conditions} Energy conservation (top) and runtime (bottom) for integrating a $1024$-particle system of fractal initial conditions for $0.25$ $N$-body units as a function of the fractal dimension. 
}
\end{figure}

\subsection{Unsoftened Plummer sphere test}\label{sec:plum}

We proceed by evolving an equal-mass $1024$-body Plummer sphere without softening through core collapse. We choose initial velocities consistent with a dynamic equilibrium, as in the softened case considered previously. This setup is chosen to match a test used on a modern implementation of a fourth-order Hermite scheme with block time steps in \cite[their section 3.4.1]{Konstantinidis:2010hx}.

In addition to the HOLD and CC schemes that have been introduced previously, we also test a modification of the CC scheme with a dedicated Kepler solver (CC\_KEPLER). In this scheme a Kepler solver is used for evolving connected components consisting of two particles. This is a form of algorithmic regularization of binaries, but note that the regularization follows naturally from the structure of the integrator and no separate binary detection or additional free parameters are necessary. The implementation of the Kepler solver is based on a universal variable formulation \citep{bate1971fundamentals}.

Results of the core collapse simulation are visualised in Figure \ref{fig:plummer-core-collapse}. All three methods produce solutions that are realistic in terms of the evolution of the mass distribution. Energy conservation is comparable to what is observed in \cite{Konstantinidis:2010hx}. Other integrals of motion show conservation around machine precision with the exception of a jump in the HOLD method around core collapse (this is caused by a high speed particle escaping from the system, causing a loss of precision in the force evaluations).

Before core collapse, execution times are roughly equivalent to the softened case considered previously (section~\ref{sec:smoothplum}) --- CC shows a modest improvement over HOLD, and CC\_KEPLER is very close to CC. Around core collapse, execution times of the HOLD and CC methods gradually increase by an order of magnitude (the CC method still consistently outperforms the HOLD method). In contrast, execution time used by the CC\_KEPLER method remains relatively uniform throughout the simulation, including core collapse.

The Sakura integrator achieves a similarly efficient treatment of close binaries by decomposing the evolution of an $N$-body Hamiltonian into a sequence of Kepler problems \citep{GoncalvesFerrari:2014uk}. The main source of errors in Sakura comes from many-body close encounters, as these are difficult to decompose into two-body interactions. In contrast, CC\_KEPLER only uses the binary solver for an isolated binary system, and switches to the regular many-body integrator when necessary (this is further discussed in Section \ref{sec:cc-ext}).

\subsection{Fractal distributions}\label{sec:fractaltest}

Since our new methods are based on the partitioning of the particle distribution in connected subsystems, we expect the method to be especially well suited to situations where substructure with extreme density contrasts exist. We therefore proceed by integrating a set of initial conditions developed with the aim of describing a star cluster with fractal substructure \citep{GoodwinWhitworth2004}. These initial conditions mimic the observed distribution of young stellar associations. They are parametrized by a fractal dimension: a low fractal dimension leads to an inhomogeneous (``structured'') distribution of stars whereas a high fractal dimension leads to a more homogenous  (``spherical'') distribution (Figure \ref{fig:time-step-graph}). For the highest possible fractal dimension value of $3$, the initial conditions approximate a constant density sphere. We use $\eta=0.03$, and integrate a $1024$-particle system under an unsoftened potential for $0.25$ $N$-body units for varying fractal dimensions.

In Figure \ref{fig:fractal-initial-conditions} we plot the energy error and runtime of  the simulation, averaged over 10 runs, as a function of the fractal dimension $f$. While all integrators show similar energy conservation, CC and CC\_KEPLER consistently outperform BLOCK and HOLD irrespective of fractal dimensions in terms of runtime. Further, runtime increases for decreasing fractal dimension for the BLOCK and HOLD integrators, while runtime remains essentially flat (and even decreases slightly) for decreasing fractal dimension for the CC and CC\_KEPLER integrators.

\subsection{Plummer sphere with binaries}\label{sec:binaries}

We proceed by looking at how our methods perform on systems containing a large number of binaries. Specifically, we take a Plummer sphere and replace every particle with a binary system. The positions and velocities of the particles are chosen such that under the absence of external perturbations, they would form a stable binary with a randomly oriented orbital plane, and a semi-major axis drawn uniformly in log space between $log(a)$ and $-0.5$. We integrate a system of $512$ binaries (=$1024$ individual particles) for $0.25$ $N$-body units with $\eta=0.03$.

\begin{figure}
\centering
\includegraphics[width=0.49\textwidth]{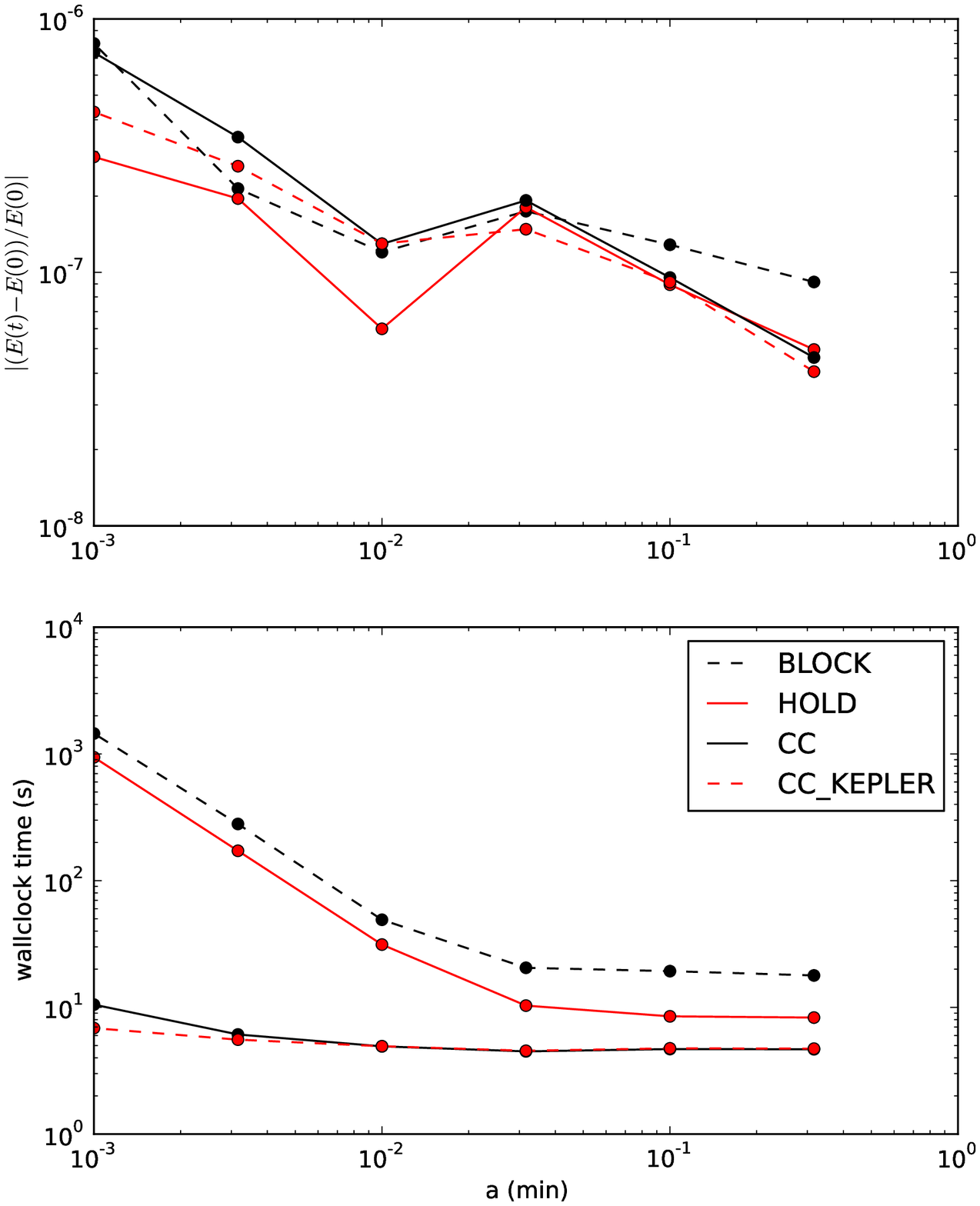}
\caption{\label{fig:plummer-of-binaries-scaling}Energy conservation (top) and runtime (bottom) for integrating $512$-binary (=$1024$-particle) Plummer sphere for $0.25$ $N$-body units as a function of the initial semi-major axis $a$.}
\end{figure}

Figure \ref{fig:plummer-of-binaries-scaling} visualises energy conservation and runtime of the initial conditions as a function of minimum semi-major axis $a$. For large $a$, the introduced binaries are generally unbounded, and the results are equivalent to evolving an ordinary Plummer sphere. As the minimum $a$ decreases, the introduced binaries become bounded and their interactions start dominating in the integration time, leading to a significant advantage for CC and CC\_KEPLER methods.

\subsection{Cold collapse test}\label{sec:coldcollapse}

As a final test we evaluate the performance of our integrators in a cold collapse scenario. Specifically, we use the fractal initial conditions described in Section \ref{sec:fractaltest} with the initial velocities set to zero. We consider a ``structured'' case with the fractal dimension $f_d=1.6$, and a ``spherical'' case with $f_d=3.0$. We evolve initial conditions for 2 $N$-body time units. For the spherical case, this is past the moment of collapse that occurs around $1.5$ $N$-body time units. For the structured case the moment of collapse is less well-defined, as different substructures collapse at different times.

We compare CC and CC\_KEPLER to two recent $N$-body codes, Ph4 (McMillan, in preparation) and HiGPUs \citep{CapuzzoDolcetta2013}. Both codes use a Hermite scheme with conventional block time steps. Ph4 implements a fourth-order scheme with the option of using the GPU-accelerated SAPORRO library \citep[][and B\'{e}dorf et al. in prep.]{Gaburov2009}. HiGPUs implements a sixth-order scheme and requires a GPU to run. We conduct our tests on a workstation --- running on a single core of an Intel i7-2720QM CPU, and a GTX460M GPU. The FLOPS performance of the GPU is roughly 40 times larger than a single core of the CPU. The hardware setup is thus indicative only of the intrinsic algorithmic scaling, rather than representative of the performance in production simulations (which would use multiple and/or more powerful CPUs/GPUs). We use unsoftened potential for CC, CC\_KEPLER, and Ph4 without GPU acceleration. We use a very small softening parameter ($\varepsilon=10^{-4}$) for Ph4 with GPU acceleration, and HiGPUS, as these run into severe slowdowns and/or crashes with unsoftened gravity --- probably because of the limited precision of their GPU kernels. We set code-specific time step accuracy parameters to $\eta=0.01$ for CC/CC\_KEPLER, $\eta_4=0.1$ for Ph4, and $\eta_4=0.05$ and $\eta_6=0.6$ for HiGPUs.

\begin{figure*}
\centering
\includegraphics[width=0.32\textwidth]{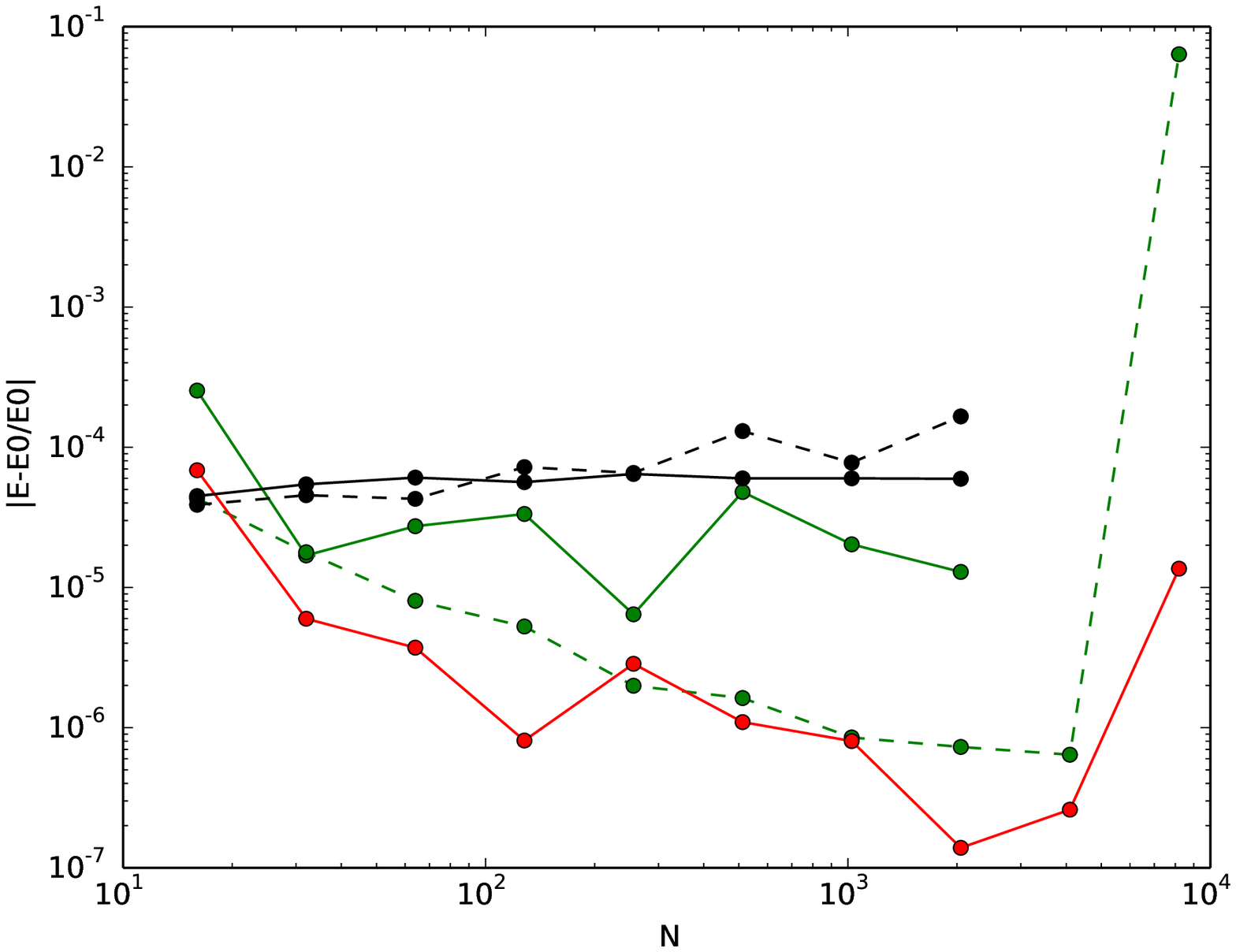}
\includegraphics[width=0.32\textwidth]{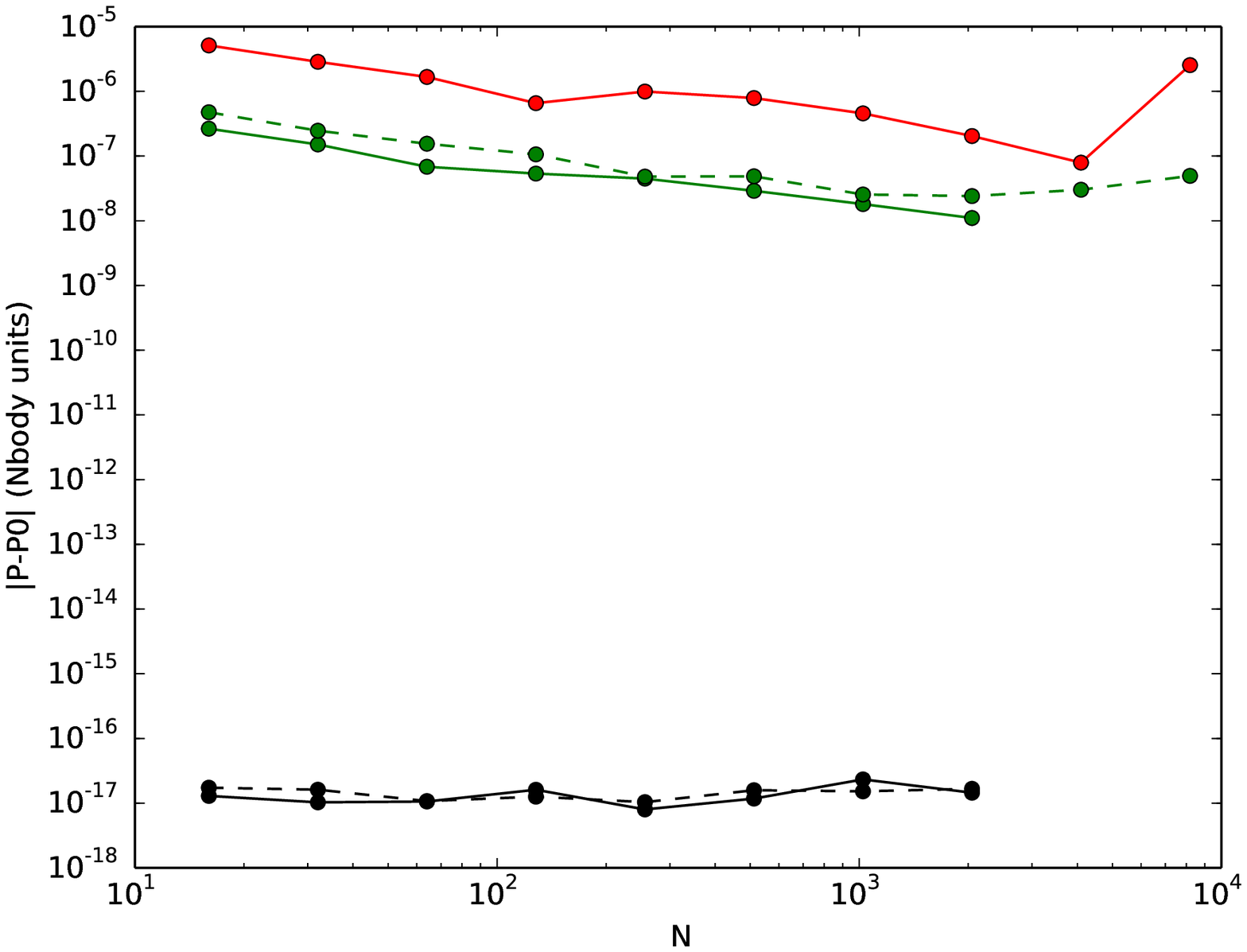}
\includegraphics[width=0.32\textwidth]{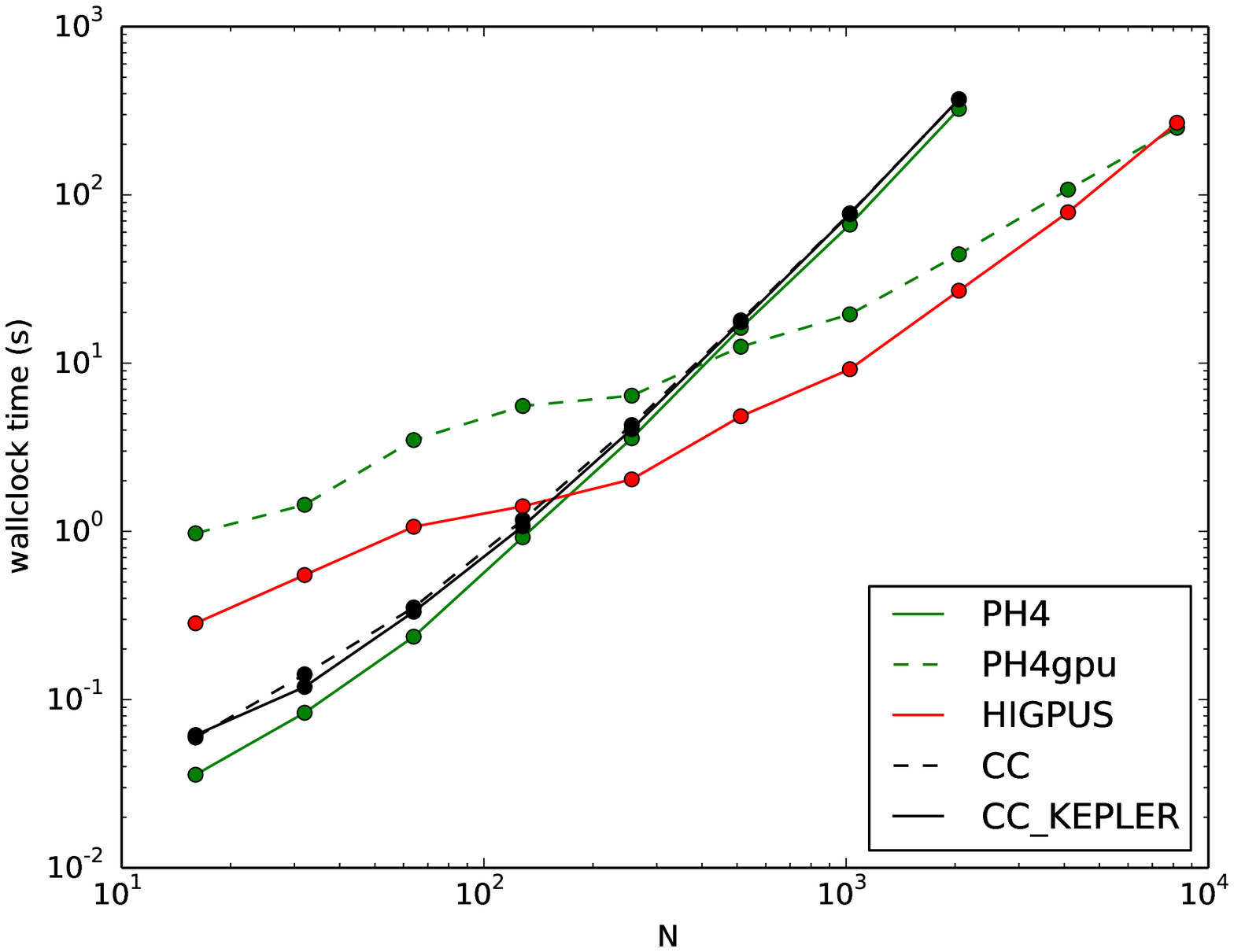}
\includegraphics[width=0.32\textwidth]{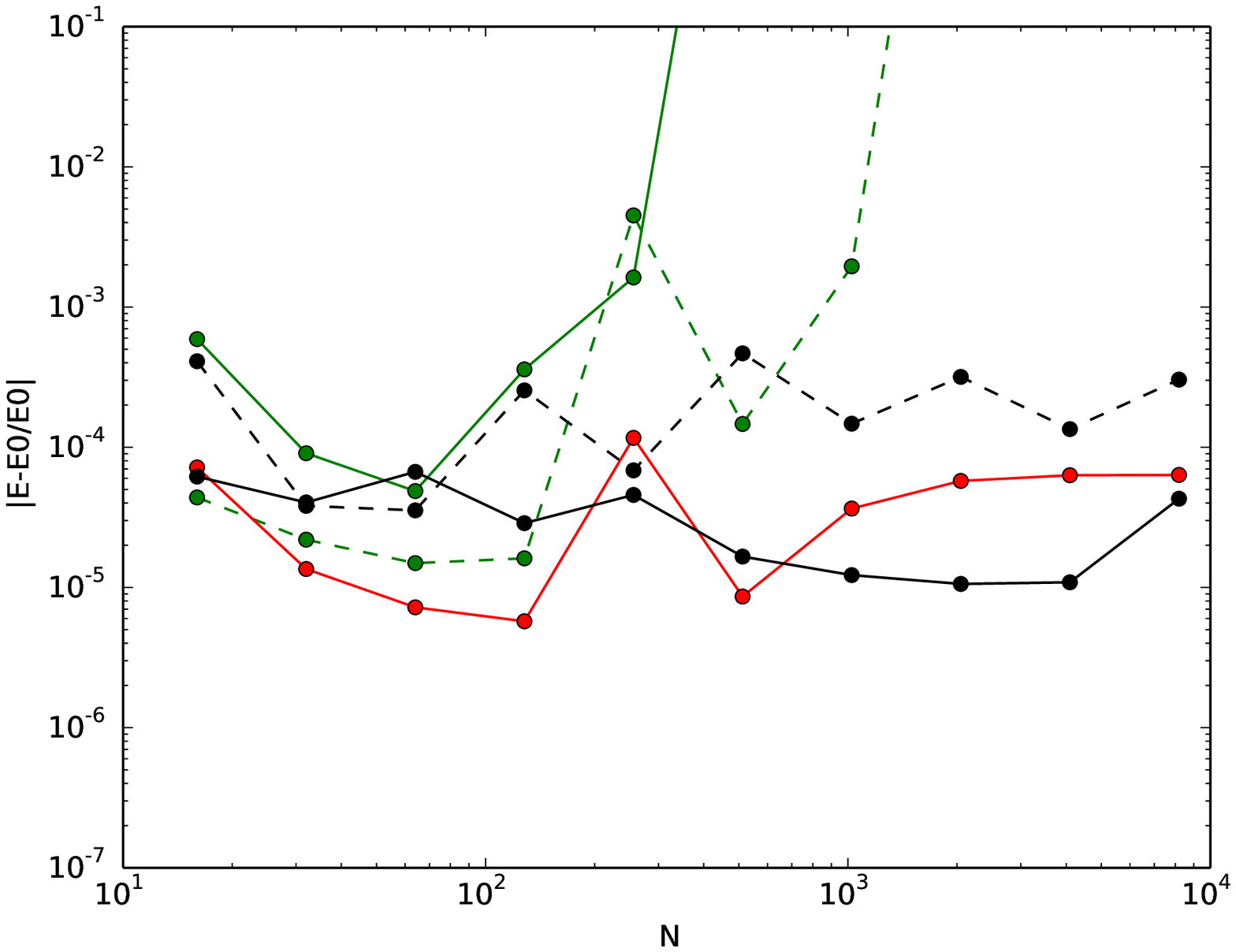}
\includegraphics[width=0.32\textwidth]{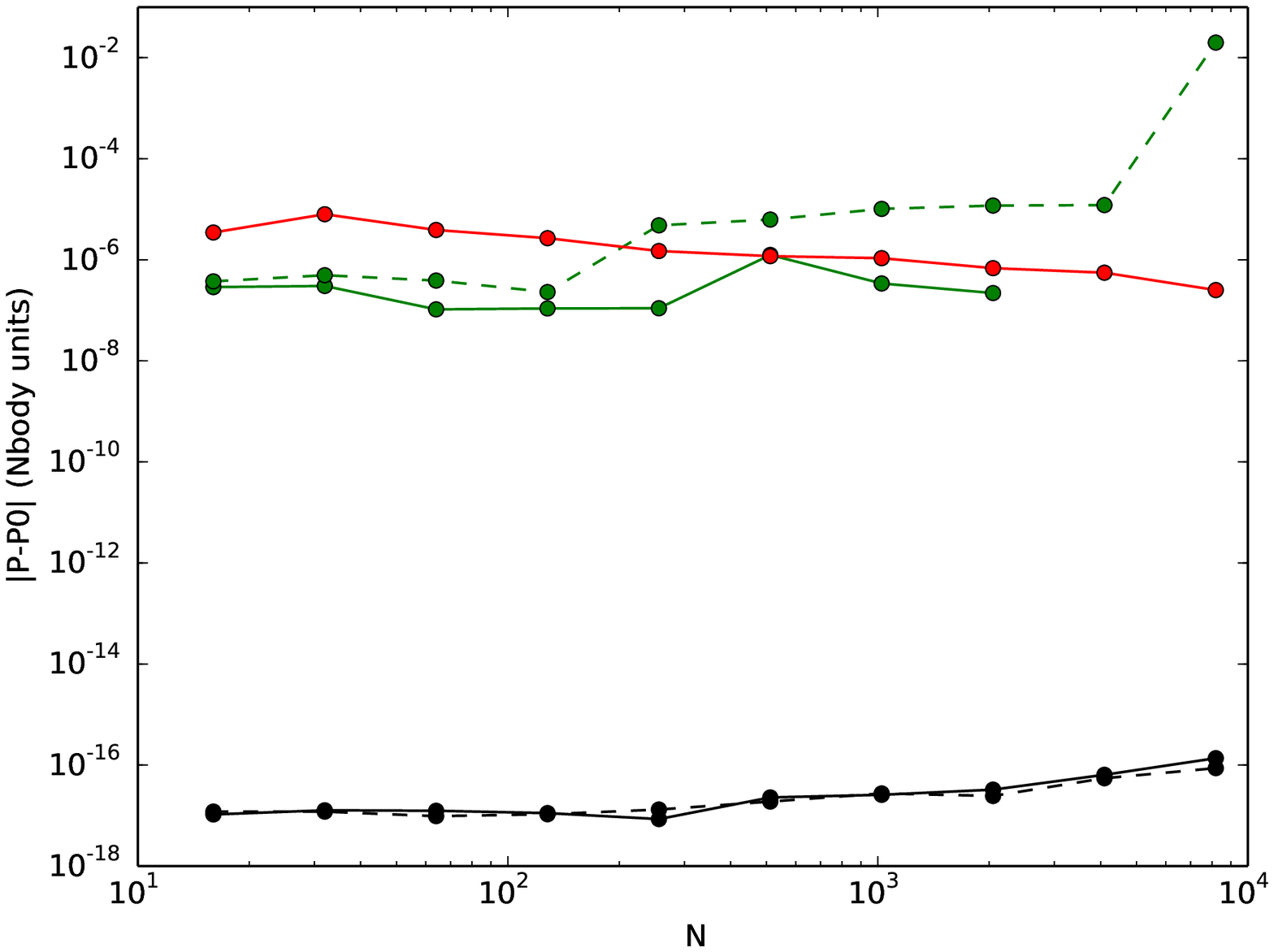}
\includegraphics[width=0.32\textwidth]{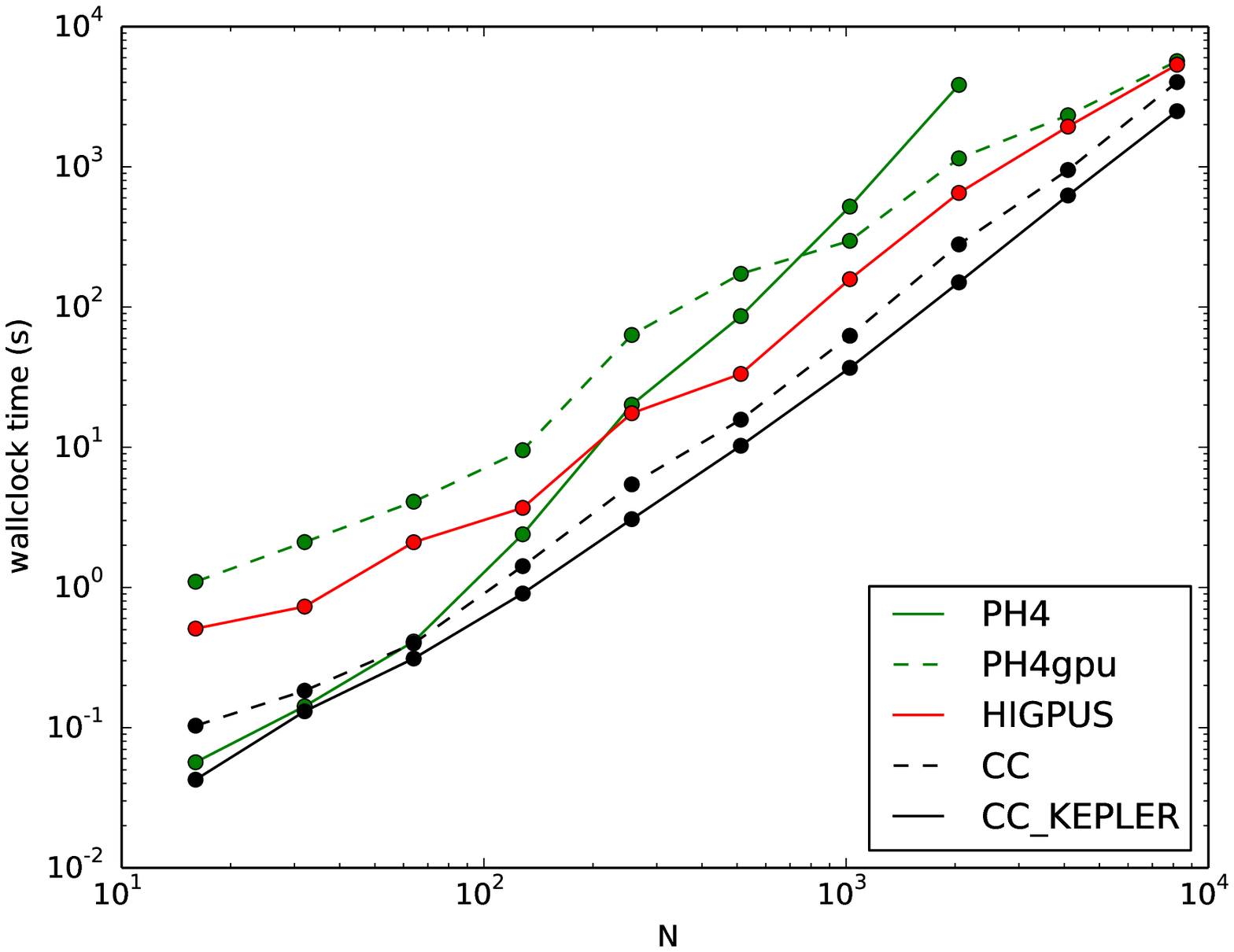}
\caption{\label{fig:cold-collapse-scaling}
Energy conservation (left column), momentum (middle column) and wall-clock time (right column) as a function of system size $N$ for the cold collapse test. The top row shows results for a homogenous sphere (fractal dimension $f_d=3$), while the bottom row shows a highly structured fractal ($f_d=1.6$). We evolve the initial conditions for 2 $N$-body time units, and plot the mean values across 5 runs. CC, CC\_KEPLER and Ph4 use unsoftened gravity ($\varepsilon=0$) , while PH4\_gpu and HIGPUS use very small softening ($\varepsilon=10^{-4}$).}
\end{figure*}

Figure \ref{fig:cold-collapse-scaling} visualises energy conservation, momentum conservation and the wall-clock time of the initial conditions as a function of the system size for structured and homogenous initial conditions. In the spherical case, setups that take advantage of the GPU (PH4\_gpu and HiGPUs) outperform the alternatives, but note that both CC and CC\_KEPLER show very similar scaling to Ph4 without GPU acceleration. In contrast, for structured case, CC\_KEPLER and CC show a marked speed up in comparison with the conventional block time step schemes, being faster for this particular calculation than Ph4\_GPU and HiGPUs, despite the latter having the advantage of using the GPU acceleration and integrating with softened gravity. The differences between the structured and the spherical cases highlight the relative advantage that the connected component approach has with respect to conventional block time steps when applied to multi-scale initial conditions.

\section{Discussion}\label{sec:discussion}

\subsection{Using and extending the CC method}\label{sec:cc-ext}

We introduced a novel method for direct integration of $N$-body systems based on splitting the system Hamiltonian using connected components of the time step graph (the CC split). We were motivated by the need for a more efficient divide-and-conquer strategy for reducing the intractable Hamiltonian (Eq.~\ref{eq:Hamiltonian}) to the least possible number of analytically solvable Hamiltonians (Eqs.~\ref{eq:kick},\ref{eq:drift}). In comparison to existing splitting methods, notably the HOLD split introduced in \cite{Pelupessy:2012if}, the CC split is particularly effective at splitting multi-scale systems. We have not encountered a situation where the HOLD split would be preferable over the CC split. The practical advantages of Hamiltonian splitting are similar to what is usually achieved with block time-steps. However, as our splitting methods, including the CC method, do not extrapolate particle states for evaluating the total force acting on a particle, we conserve linear and angular momentum to machine precision. 

We went on to show on the example of the CC\_KEPLER method that the connected components partitioning has additional uses beyond improved splitting efficiency. Specifically, we were able to incorporate regularization of two-body close encounters by simply checking for the condition where the successive partitioning leads to a connected component with two particles, and evolving the corresponding two-body Hamiltonian (Eq \ref{eq:kepler}) using a dedicated Kepler solver. This approach can be extended to many-body close encounters by using a suitable specialised solver \citep[or e.g. chain regularization methods,][]{Mikkola2008} to evolve isolated Hamiltonians corresponding to connected components with certain properties. Possible selection criteria include having a specific number of particles and/or a maximum time step below a threshold value or the structure of the timestep graph. 

The numerical experiments of Section \ref{sec:tests} were chosen to mainly study the splitting aspect of $N$-body integration. We focused on normalised performance metrics, and the scaling of the wall-clock time as a function of the ``multi-scaleness'' in the initial conditions. Our current implementations would benefit from additional optimisations typically used in production-level $N$-body codes. Specifically, there is inherent parallelism in the CC method, as recursive calls for evolving successively smaller closed Hamiltonians only affect the state of the particles in the ``current'' closed component. It may be possible to parallelise the method based on this property. However, tests show that a naive approach does not scale well due to load-balancing issues, as subsystems can vary substantially in size. Alternatively, it could be feasible to implement the CC method on a GPU, as
the major components --- $N$-body force evaluation\citep{PortegiesZwart:2007gl, Belleman:2008fv, CapuzzoDolcetta2013} and graph processing algorithms \citep{Harish07} --- have individually been successfully implemented on GPUs.

It may be possible to speed up the evaluation of long-range interactions between different connected components ($V_{CC}$ in the CC decomposition formula) through a centre-of-mass (or multipole) approximation that form the basis of tree codes \cite{Barnes:1986ed}. As long-range interactions between two connected components are evaluated symmetrically, this approach could make it possible to obtain most of the speedup of a tree code while maintaining good linear and angular momentum conservation. A potential pitfall with this approach could arise from the fact that the time step criterion $\tau(i,j)$ used in finding the connected components is only partially determined by the coordinates of the particles. As such, particles in the same connected component may occupy a ``non-compact'' region in physical space, making multipole approximation difficult.

\subsection{Formally optimal Hamiltonian splitting}

The HOLD integrator determines the accuracy of a kick between particles $i$ and $j$ from the particle-based time steps $\tau(i)$ and $\tau(j)$. In the CC integrator the accuracy of a kick is determined by the time step graph generated directly from interaction-based time steps $\tau(i,j)$. Could we further improve the splitting by applying kicks directly based on the interaction-specific time step criteria $\tau(i,j)$?

We implemented this idea in an experimental integrator which we named the OK split (OK stands for \emph{Optimal Kick}). The method partitions a list of all {\em interactions} in the system (based on a pivot time step $h$) just like the HOLD split partitions a list of all {\em particles} in the system. The partitioning is formally optimal in the sense that every kick is evaluated at the time step closest to $\tau(i,j)$ in the power of two hierarchy based on the pivot time step $h$. While the possibility of direct $N$-body integration with interaction-based time steps has been previously considered in \cite{Nitadori:2008gt}, the OK split is the first workable implementation of this idea that we are aware of.

In numerical tests, the OK split is not competitive compared to other methods such as the CC split. For example, in the $1024$-body smoothed Plummer sphere test from Section \ref{sec:smoothplum}, the relative energy error at the end of the simulation is around $10^{-2}$ (several orders of magnitude worse than HOLD and CC, but possibly still enough for drawing statistically correct conclusions, \cite{PortegiesZwart2014}). The remaining integrals of motion are conserved at machine precision, as the OK split applies kicks in pairs. Finally, the evolution of the mass distribution is comparable to HOLD and CC with the OK split using fewer kick and time step evaluations.

Could we improve the OK split by changing the time step criteria? For example, consider $\tau_{\star}(i,j)=\min\left(\tau(i),\tau(j)\right)$ where $\tau$ is the particle-based time step criteria as defined in \cite{Pelupessy:2012if}. Formally, combining the OK split with $\tau_{\star}(i,j)$ would result in a splitting with the exact same kicks and drifts as the HOLD integrator. This somewhat contrived example only serves the point of illustrating that the time step criteria can qualitatively change the behaviour of the OK split. While it is unknown whether practical interaction-based time step criteria even exist, we do believe that a closer look at the various simplifications made during the derivation of the explicit and approximately time-symmetric time step criteria that we've used throughout this work (eq \ref{eq:tauij}) would serve as a good starting point.

\begin{acknowledgements}
We thank Guilherme Gon\c{c}alves Ferrari and the anonymous referee
for a critical reading of the manuscript. This work was supported by the Netherlands Research Council NWO (Grants \#643.200.503, \#639.073.803 and \#614.061.608) and by the Netherlands Research School for Astronomy (NOVA). J\"urgen J\"anes was supported by the Archimedes Foundation, Estonian Students' Fund USA, Estonian Information Technology Foundation and Skype.
\end{acknowledgements}

\bibliographystyle{aa}
\bibliography{cc_paper}
\end{document}